\documentstyle[12pt,aaspp4,epsfig]{article}

\def\ls{{_<\atop^{\sim}}}
\def\gs{{_>\atop^{\sim}}}
\def\cgs{{ erg cm$^{-2}$ s$^{-1}$}}

\begin{document}

\title{A flash in the dark: UVES/VLT high resolution spectroscopy
of GRB afterglows\footnote{Based on observations collected at the European
Southern Observatory, ESO, the VLT/Kueyen telescope, Paranal, Chile, in the
framework of programs 69.A--0516(B) and 70.A--0599(B)}} 

\author{F. Fiore$^1$, V. D'Elia$^1$, D. Lazzati$^{2,3}$, R. Perna$^{3,4}$,
L. Sbordone$^{1,5}$, G. Stratta$^{1,6}$, E.J.A. Meurs$^7$, P. Ward$^7$, 
L.A. Antonelli$^1$,
G. Chincarini$^8$, S. Covino$^9$, A. Di Paola$^1$, A. Fontana$^1$,
G. Ghisellini$^9$, G. Israel$^1$, F. Frontera$^{10}$, G. Marconi$^{1,5}$,
L. Stella$^1$, M. Vietri$^{11}$, F. Zerbi$^9$\\
$^1$INAF--Osservatorio Astronomico di Roma, Via Frascati 33, I--00044 
Monteporzio Catone, Italy;\\
$^2$ Institute of Astronomy, University of Cambridge, Madingley Road, 
Cambridge, UK;\\
$^3$ Department of Astrophysical and Planetary Science, CU Boulder, Boulder,
80309, USA;\\
$^4$ Princeton University Observatory, Princeton, NJ 08544--1001, USA;\\
$^5$ European Southern Observatory, Casilla 19001, Santiago, Chile;\\
$^6$ Dipartimento di Fisica, Universita' di Roma La Sapienza, Italy;\\
$^7$ Dunsink Observatory, Castleknock, Dublin 15, Ireland;\\
$^8$ Universita' di Milano Bicocca, Piazza della Scienza 3, 20126 Milano, 
Italy\\
$^9$ INAF, Osservatorio Astronomico di Brera, via E. Bianchi 46, 23807 
Merate (LC), Italy.\\
$^{10}$ Dipartimento di Fisica, Universita' di Ferrara, via Paradiso 12, 
44100 Ferrara, Italy;\\
$^{11}$ Scuola Normale Superiore, I--56100 Pisa, Italy.
}

\author{\tt (version: 2 February 2005) }

\begin{abstract}
We present the first high resolution (R=20000--45000, corresponding to
14 km/s at 4200\AA\ to 6.6 km/s at 9000\AA\ ) observations of the
optical afterglow of Gamma Ray Bursts. GRB 020813 and GRB 021004 were
observed by UVES@VLT 22.19 hours and 13.52 hours after the trigger,
respectively. These spectra show that the inter--stellar matter of the
GRB host galaxies is complex, with many components contributing to
each main absorption system, and spanning a total velocity range of up
to about 3000 km/s. Several narrow components are resolved down to a
width of a few tens of km/s. In the case of GRB 021004 we detected
both low and high ionization lines.  Combined with photoionization
results obtained with CLOUDY, the ionization parameters of the various
systems are consistent with a remarkably narrow range with no clear
trend with system velocity. This can be interpreted as due to density
fluctuations on top of a regular $R^{-2}$ wind density profile.
\end{abstract}
 
 
\section{Introduction}

For a few hours after their onset, Gamma Ray Bursts (GRBs) are the
brightest beacons in the far Universe, offering a superb opportunity
to investigate both GRB physics and high redshift galaxies.  Tens of
minutes after a GRB, its optical afterglow can be as bright as
magnitude 13--15; a few hours later a magnitude of 16--19 is often
achieved.  Bright examples are the cases of GRB 990123 (z=1.600),
for which the reverse shock reached R=9-10 mag 1 minute after the GRB
and the optical afterglow reached R=14 mag 12 minutes after the GRB
(e.g. Akerlof et al. 1999, Galama et al. 1999), and GRB 030329 (z=0.1685), 
for which the optical afterglow reached R=12.7 mag at 1.5
hours from the GRB and decreased down to R=19 mag after $\sim 10$ days
(e.g. Price et al. 2003; Stanek et al. 2003).  High resolution (a few
tens of km/s in the optical band), high quality (signal to noise $>10$
per resolution element) spectra can therefore be gathered, provided
that the afterglow is observed rapidly by 8m class telescopes.
Since GRBs are associated with the collapse of massive stars (see
Woosley 1993, Paczynski 1998, MacFadyen \& Woosley 1999, Vietri \&
Stella 1998 for theoretical reasons and Galama et al. 1998,
Stanek et al. 2003, Hjorth et al. 2003 and Della Valle et al. 2003 for
observational reasons), this is expected to open a new window in
the study of the environment in which intermediate to high redshift
star--formation occurs and, in particular, on the physical, chemical
and dynamical state of the inter--stellar matter (ISM) of the GRB host
galaxies.

The study of z$\gs1$ galaxies has so far  mostly relied upon
Lyman--break galaxies (LBGs) at z$=3--4$ (see e.g. Steidel et al.
1999) and galaxies which happen to be along the line of sight to
bright background QSOs. Some of these systems are associated with
Damped Ly$\alpha$ systems (DLA, see e.g.  Pettini et al. 1997).
However, LBGs are characterized by pronounced star--formation and
their inferred chemical abundances may be related to these regions
rather than being representative of typical high z galaxies.  
Metal line systems associated to DLAs along the line of sight to
quasars probe mainly galaxy haloes, rather than their bulges or discs.
Furthermore, it is not clear whether galaxies associated with DLAs are
truly representative of the whole high--z galaxy population.  GRB
afterglows provide an independent tool to study the ISM of high z
galaxies.  Savaglio, Fall \& Fiore (2003) studied the metal abundances
in three GRB host galaxies using low--medium resolution spectroscopy
and a curve of growth analysis. They found metal column densities
higher that in QSO--DLAs and a strong inverse correlation between
[Si/Zn], [Cr/Zn] and [Fe/Zn] and the Zn column density, indicating a
dense environment and large dust depletion.  On the other hand,
Vreeswijk et al. (2004) found indications for a relatively low
metallicity and low dust content in the ISM of the z=3.372 host galaxy
of GRB 030323, using FORS2 low and intermediate resolution
spectroscopy.  Using again a curve of growth analysis, and taking
advantage of ultra--deep {\it Gemini} multi--object spectrograph
observations, Savaglio et al. (2004) studied the ISM of a sample of
faint K band selected galaxies at 1.4$<$z$<$2.0, finding \ion{Mg}{2}
and \ion{Fe}{2} abundances much higher than in QSO--DLAs and similar
to those in GRB host galaxies.
A much better job can be done with high resolution ($R>20000$)
observations because: (i) lines can be separated; (ii) metal
column densities can be measured through a fit to the line profile;
(iii) fainter lines can be measured; (iv) information on the gas
dynamics in the GRB host galaxies can be derived.  Furthermore, as
suggested by the comparison of the Vreeswijk et al. (2004) and the
Savaglio et al. (2004) studies, GRB afterglows allow us to probe
galaxy ISM at much higher redshifts than even ultra--deep, standard
galaxy spectroscopy, such as the Gemini Deep Deep Survey.

For all these reasons we started a pilot program to observe bright GRB
afterglows of promptly localized GRBs with UVES@VLT. 
The program has been conceived and designed to make
full use of the
%
GRB afterglows discovered by the GRB--dedicated {\it Swift} mission,
launched on November 20 2004\footnote{http://swift.gsfc.nasa.gov}.
{\it Swift} will provide GRB positions ($<4\arcmin$ precision) in
$<10$~s, X--ray afterglow positions ($3\arcsec$ precision) in
$<100$~s, and an optical finding chart in $<300$~s, with a few arcsec
to sub--arcsec position accuracy, thus revolutionizing fast response
multiwavelength studies of GRBs.

This article concentrates on high resolution spectroscopy of two GRB
afterglows.  The HETE--2 FREGATE, WXM and SXC instruments detected
GRB 021004 on 2002 Oct. 4 12:06:13.57 UT (Shirasaki et al. 2002) and
GRB 020813 on 2002 Aug. 13 02:44:19.17 UT (Villasenor et al. 2002).
The WXM flight localization software produced a reliable and
relatively accurate position 49 seconds after the burst for GRB 021004
(error box of 30$\arcmin$ radius) and 4 minutes after the burst for
GRB 020813 (error box of 14$\arcmin$ radius).  Bright optical
counterparts were identified 10 minutes and 1.9 hours after the
triggers by Fox  et al. (2002, 2003),  and Fox, Blake \& Price (2002)
respectively. Ground analysis of the HETE2 data performed a few hours
after each GRB improved the localization of the event, providing error
boxes of 2$\arcmin$ and 1$\arcmin$ radius, respectively.  UVES
observations started 13.52 hours after the GRB 021004 trigger,
(Savaglio et al. 2002) and 22.19 hours after the GRB 020813 trigger
(Fiore et al.  2002).

Optical low to intermediate resolution spectroscopy of the
afterglow of GRB 021004 was obtained by Moller et al. (2003), Mirabal et
al. (2003), Matheson et al. (2003), Schaefer et al. (2003).  One of the main
conclusion of these works is that density fluctuations on top of a
regular wind density profile are able to explain both the presence of
strong, blue--shifted, high ionization absorption line systems and
irregularities in the optical light curve. Schaefer et al. (2003)
propose a scenario where a clumpy wind originates from a massive star
progenitor (a Wolf--Rayet star), while Mirabal et al. (2003) suggest
that the ionized absorption takes place in a fragmented shell nebula
around the GRB progenitor.  Optical spectropolarimetry of the
afterglow of GRB 020813 has been presented by Barth et al. (2003) who
report the detection of strong absorption systems at z=1.223 and
z=1.255. A further analysis of the Keck LRIS spectrum
is presented by Savaglio et al. (2004). 
The present paper reports the results of higher resolution
spectroscopy of both afterglows.

Both GRBs were observed in the X--rays too by the Chandra High Energy
Transmission Grating System, which has a resolution of $\sim 1000$ at
1 keV (Butler et al. 2003 and references therein). 
Both observations lasted about 80 ks and started 21 hours and
20 hours after the GRB event for GRB 020813 and GRB 021004 respectively.
Fading X--ray afterglows were detected in both cases with decay indices
of $-1.38\pm0.06$ and $-0.9\pm0.1$, consistent with the values
reported for the optical afterglows by Covino et al. (2003) and
Holland et al. (2003). The time averaged fluxes were
$2.2\times10^{-12}$ \cgs and $6.3\times10^{-13}$\cgs respectively. The
X--ray spectra are consistent with power laws of energy index of
$-0.85\pm0.04$ and $-1.01\pm0.08$ reduced at low energy by Galactic
column density (which is rather high in the direction of GRB 020813,
$N_{HGal}=7.5\times10^{20}$ cm$^{-2}$, and is
$N_{HGal}=4.2\times10^{20}$ cm$^{-2}$ in the direction of GRB 021004).
The upper limits to the rest frame X--ray absorbing column are of the
order of a few$\times10^{20}$ in both cases.  Butler et al. (2003)
report the detection of an emission line at 1.3 keV at the 3.3$\sigma$
confidence level, interpreted as a SXVI l$\alpha$ line blue--shifted by
0.12c.

\section{Observations and data reduction}

In the framework of ESO programs 69.A--0516 and 70.A--0599 we observed
the afterglows of GRB 020813 and GRB 021004 with the high resolution
UV--visual echelle spectrograph (UVES, Dekker et al. 2000), mounted at
the VLT--UT2 telescope.  Table 1 gives the log of the observations
used in this paper. A further 1 hour dichroic 1 observation is
available in the ESO archive, but we were not able to extract a
reliable spectrum from this observation, which has therefore been
discarded.  In order to maximize the signal to noise ratio the CCD was
rebinned $2\times2$ pixels. The data reduction has been performed
using the UVES
pipeline\footnote{http://www.eso.org/observing/dfo/quality/UVES/pipeline/}.
The final useful spectra extend from about 4250\AA\ to about 9400\AA\
and were rebinned to 0.1\AA\ to increase the signal to noise. The
resulting resolution element, set to two pixels, ranges then from 14
km/s at 4200\AA\ to 6.6 km/s at 9000\AA\ .  The noise spectrum, used
to determine the errors on the best fit line parameters, has been
calculated from the real background subtracted and rebinned spectrum
using line free regions. This therefore takes into account both
statistical errors and systematic errors in the pipeline processing
and background subtraction. Table 2 gives the signal to noise ratio
per 0.1\AA\ pixel at different wavelengths for the 4 spectra in Table
1.  Figure 1 plots the full UVES spectrum of GRB 021004, for which both
UVES dichroics, and both red and blue arms, were used, allowing us to
obtain a particularly wide spectral coverage, extending from
$\sim$4250\AA\ to $\sim$9400\AA\ .  A further dichroic 1, blue arm
image, covering the band 3400\AA\ -- 3900 \AA\ was obtained, but the
signal to noise is too low to allow a reliable extraction of the
spectrum in this band. Figure \ref{spe020813} plots the full UVES
spectrum of GRB 020813. This spectrum, as well as the spectrum in
figure 1,  was smoothed with a gaussian function with $\sigma=1.5$
pixels. Tables 3 and 4 give the equivalent width of all lines detected
in the UVES spectra for the two GRB afterglows, along with their
identification. The equivalent widths are computed at the
redshift of the absorption systems. For the GRB host galaxy identified
systems and the 2 main intervening systems at z=1.60 and 1.38 we
report also faint lines with (signal to noise $>1$). Unidentified
lines are reported only if the signal to noise is $>3$. Their
equivalent widths are computed at zero redshift.

\begin{table}
\caption{\bf Journal of observations}
\footnotesize
\begin{tabular}{lccccccc}
\tableline\tableline
Date UT  & Dichroic & B. Arm C.W. & R. Arm C.W. & slit$^{a}$ 
& seeing$^{a}$ & exposure$^{b}$ & time since GRB$^{c}$\\
\hline
14/08/02 00:55:25 & 1	& 3460\AA\ & 5800\AA\ & 1 & $\ls1$ & 60 & 22.19 \\
05/10/02 01:37:37 & 2	& 4370\AA\ & 8600\AA\ & 1 & $\ls1$ & 30 & 13.52 \\
05/10/02 02:10:53 & 1	& 3460\AA\ & 5800\AA\ & 1 & $\ls1$ & 30 & 14.08 \\
05/10/02 04:09:54 & 1	& 3460\AA\ & 5800\AA\ & 1 & $\ls1$ & 60 & 16.06 \\
\tableline
\end{tabular}
\normalsize

$^{a}$arcsec;
$^{b}$min;
$^{c}$hr
\end{table}

\begin{table}
\caption{\bf Signal to noise ratio}
\begin{tabular}{lccc}
\tableline\tableline
$\lambda$	& 14/08/02 00hr & 05/10/02 01hr-02hr & 05/10/02 04hr \\
\hline
4250\AA\	&	& 3.6	&	\\	
5000\AA\	& 4.7	& 5.9	& 9.3	\\	
6000\AA\	& 5.6	& 5.4	& 11.2	\\	
7000\AA\	& 5.9	& 9.8	& 10.2	\\	
7800\AA\	&	& 6.8	&	\\	
9250\AA\	&	& 2.9	&	\\	
\tableline
\end{tabular}
%

\end{table}

\begin{table}
\caption{\bf GRB 021004 UVES line identifications}
\footnotesize
\begin{tabular}{lccc}
\hline
\hline
$\lambda$(\AA\ ) & $W_{rest}$(\AA\ )$^a$&   ID  &    z  \\
\hline
4298.40   &  0.13$\pm$0.02   & ?\ion{Si}{2}$\lambda$1304.37?   & ?2.2953?  \\
4301.73   &  0.06$\pm$0.03   &  \ion{Si}{2}$\lambda$1808.01    &  1.3793-8 \\
4304.53   &  0.09$\pm$0.03   &  \ion{Si}{2}$\lambda$1808.01    &  1.3807  \\
4311.20   &  0.16$\pm$0.08   &                    &          \\
4346.30   &  0.27$\pm$0.08   &  \ion{Al}{2}$\lambda$1670.79    &  1.6014  \\
4347.25   &  0.18$\pm$0.03   &  \ion{Al}{2}$\lambda$1670.79    &  1.6019  \\
4348.60   &  0.11$\pm$0.03   &  \ion{Al}{2}$\lambda$1670.79    &  1.6028  \\
4350.95   &  0.53$\pm$0.08   &                    &          \\
4358.28   &  0.35$\pm$0.08   &                    &          \\
4398.34   &  0.06$\pm$0.03   &  \ion{C}{2}$\lambda$1334.53     & 2.2958 \\
4401.68   &  0.24$\pm$0.02   &  \ion{C}{2}$\lambda$1334.53     & 2.2981-4 + \\
          &                  & ?\ion{C}{2}$^*\lambda$1335.70?  & ?2.2958?  \\
4415.50   &  0.10$\pm$0.03   &  \ion{Al}{3}$\lambda$1854.71    &  1.3807  \\
4441.34   &  0.04$\pm$0.02   &  \ion{C}{2}$\lambda$1334.53     &  2.328 \\
4445.35   &  0.09$\pm$0.02   &  \ion{C}{2}$^*\lambda$1335.70   &  2.328 \\
4593.59   &  0.15$\pm$0.02   &  \ion{Si}{4}$\lambda$1393.76    &  2.2958  \\
4596.71   &  0.02$\pm$0.02   &  \ion{Si}{4}$\lambda$1393.76    &  2.2981-4\\
4623.48   &  0.21$\pm$0.02   &  \ion{Si}{4}$\lambda$1402.77    &  2.2958  \\
4626.41   &  0.02$\pm$0.02   &  \ion{Si}{4}$\lambda$1402.77    &  2.2981-4\\
4629.72   &  0.40$\pm$0.02   &  \ion{Si}{4}$\lambda$1393.76    &  2.321   \\
4631.35   &  0.30$\pm$0.07   &                    &          \\
4635.85   &  0.65$\pm$0.07   &                    &          \\
4637.38   &  3.63$\pm$0.07   &                    &          \\
4638.01   &  0.31$\pm$0.02   &  \ion{Si}{4}$\lambda$1393.76    &  2.328   \\
4643.19   &  0.25$\pm$0.07   &                    &          \\
4658.07   &  0.27$\pm$0.02   &  \ion{Si}{4}$\lambda$1402.77    &  2.321   \\
4666.28   &  1.07$\pm$0.07   &                    &          \\
4668.38   &  0.56$\pm$0.02   &  \ion{Si}{4}$\lambda$1393.76    &  2.328   \\
5102.54   &  0.50$\pm$0.02   &  \ion{C}{4}$\lambda$1548.20     &  2.2958  \\
5106.50   &  0.35$\pm$0.02   &  \ion{C}{4}$\lambda$1548.20     &  2.2981-4\\
5111.04   &  0.33$\pm$0.02   &  \ion{C}{4}$\lambda$1550.77     &  2.2958  \\
5114.82   &  0.16$\pm$0.02   &  \ion{C}{4}$\lambda$1550.77     &  2.2981-4\\
5141.11   &  1.74$\pm$0.02   &  \ion{C}{4}$\lambda$1448.20     &  2.321   \\
5150.4    &  2.10$\pm$0.02   &  \ion{C}{4}$\lambda$1550.77     &  2.321 + \\
          &                  &  \ion{C}{4}$\lambda$1448.20     &  2.328   \\
5159.66   &  1.57$\pm$0.02   &  \ion{C}{4}$\lambda$1550.77     &  2.328   \\
5510.62   &  0.05$\pm$0.02   &  \ion{Al}{2}$\lambda$1670.79    &  2.2981-4\\
5560.30   &  0.47$\pm$0.02   &  \ion{Al}{2}$\lambda$1670.79    &  2.328   \\
5577.71   &  1.47$\pm$0.03   &  \ion{Fe}{2}$\lambda$2344.21    &  1.3793-8\\
5580.81   &  0.07$\pm$0.03   &  \ion{Fe}{2}$\lambda$2344.21    &  1.3807  \\
5592.94   &  0.14$\pm$0.06   &                    &          \\
5649.59   &  0.31$\pm$0.03   &  \ion{Fe}{2}$\lambda$2374.21    &  1.3793-8\\
5669.68   &  0.67$\pm$0.03   &  \ion{Fe}{2}$\lambda$2382.77    &  1.3793-8\\
6064.56   &  0.60$\pm$0.06   &                    &          \\
6098.32   &  0.22$\pm$0.06   &                    &          \\
\hline
\end{tabular}
\normalsize

Errors are 67\% confidence intervals.
$^a$ the equivalent width of unidentified lines is computed at
zero redshift;
? = uncertain identification. In many cases several components contribute to 
a single line in one entry of the table. In these cases either the redshift
is given with only 4 digits or it is given as a range. 
A + sign indicates that
a line is partly blended with the following entry.
\end{table}

\setcounter{table}{2}

\begin{table}
\caption{\bf GRB 021004 UVES line identifications, continued.}
\footnotesize
\begin{tabular}{lccc}
\hline
\hline
$\lambda$(\AA\ ) & $W_{rest}$(\AA\ )$^a$&   ID  &    z  \\
\hline
6099.33   &  0.19$\pm$0.02   &  \ion{Fe}{2}$\lambda$2344.21    &  1.6019  \\
6101.32   &  0.09$\pm$0.02   &  \ion{Fe}{2}$\lambda$2344.21    &  1.6028  \\
6154.64   &  0.52$\pm$0.03   &  \ion{Fe}{2}$\lambda$2586.65    &  1.3793-8\\
6158.03   &  0.01$\pm$0.03   &  \ion{Fe}{2}$\lambda$2586.65    &  1.3807  \\
6160.76   &  0.17$\pm$0.06   &                    &          \\
6172.42   &  0.35$\pm$0.02   &  \ion{Al}{3}$\lambda$1854.72    &  2.328   \\
6176.74   &  0.21$\pm$0.06   &                    &          \\
6177.88   &  0.17$\pm$0.02   &  \ion{Fe}{2}$\lambda$2374.46    &  1.6019  \\
6180.04   &  0.14$\pm$0.05   &                    &          \\
6186.51   &  0.25$\pm$0.02   &  \ion{Al}{3}$\lambda$1862.79    &  2.321 + \\   
6186.99   &  0.11$\pm$0.02   &  \ion{Fe}{2}$\lambda$2600.17    &  1.3793-8\\  
6187.81   &  0.05$\pm$0.02   &  \ion{Fe}{2}$\lambda$2600.17    &  1.3798  \\  
6190.23   &  0.12$\pm$0.02   &  \ion{Fe}{2}$\lambda$2600.17    &  1.3807  \\
6199.36   &  0.11$\pm$0.02   &  \ion{Al}{3}$\lambda$1862.79    &  2.328 +  \\
6198.4    &                  &                                 &           \\
6199.53   &  0.29$\pm$0.02   &  \ion{Fe}{2}$\lambda$2382.77    &  1.6019  \\
6201.66   &  0.28$\pm$0.02   &  \ion{Fe}{2}$\lambda$2382.77    &  1.6028  \\
6653.88   &  0.89$\pm$0.02   &  \ion{Mg}{2}$\lambda$2796.35    &  1.3793-8\\
6655.88   &  0.24$\pm$0.02   &  \ion{Mg}{2}$\lambda$2796.35    &  1.3802  \\
6657.60   &  0.47$\pm$0.02   &  \ion{Mg}{2}$\lambda$2796.35    &  1.3807  \\
6671.00   &  0.78$\pm$0.02   &  \ion{Mg}{2}$\lambda$2803.53    &  1.3793-8\\
6672.73   &  0.06$\pm$0.02   &  \ion{Mg}{2}$\lambda$2803.53    &  1.3802  \\
6674.77   &  0.32$\pm$0.02   &  \ion{Mg}{2}$\lambda$2803.53    &  1.3807  \\
6730.16   &  0.25$\pm$0.02   &  \ion{Fe}{2}$\lambda$2586.65    &  1.6019  \\
6732.55   &  0.09$\pm$0.02   &  \ion{Fe}{2}$\lambda$2586.65    &  1.6028  \\
6750.64   &  0.08$\pm$0.02   &  \ion{Mn}{2}$\lambda$2594.50    &  1.6019  \\
6764.18   &  0.43$\pm$0.04   &                    &          \\
6765.43   &  0.25$\pm$0.02   &  \ion{Fe}{2}$\lambda$2600.17    &  1.6019  \\
6767.79   &  0.17$\pm$0.02   &  \ion{Fe}{2}$\lambda$2600.17    &  1.6028  \\
6768.78   &  0.18$\pm$0.04   &                    &          \\
6781.81   &  0.04$\pm$0.02   &  \ion{Mn}{2}$\lambda$2606.46    &  1.6019  \\
6788.06   &  0.03$\pm$0.02   &  \ion{Mg}{1}$\lambda$2852.94    &  1.3793-8\\
7003.91   &  0.12$\pm$0.03   &                    &          \\
7041.91   &  0.12$\pm$0.03   &                    &          \\
7042.52   &  0.11$\pm$0.03   &                    &          \\
7043.25   &  0.10$\pm$0.03   &                    &          \\
7274.52   &  0.26$\pm$0.01   &  \ion{Mg}{2}$\lambda$2796.35    &  1.6014  \\
7275.83   &  0.33$\pm$0.01   &  \ion{Mg}{2}$\lambda$2796.35    &  1.6019  \\
7277.28   &  0.49$\pm$0.01   &  \ion{Mg}{2}$\lambda$2796.35    &  1.6024  \\
7278.39   &  0.24$\pm$0.01   &  \ion{Mg}{2}$\lambda$2796.35    &  1.6028  \\
7293.12   &  0.27$\pm$0.01   &  \ion{Mg}{2}$\lambda$2803.35    &  1.6014  \\
7294.58   &  0.36$\pm$0.01   &  \ion{Mg}{2}$\lambda$2803.53    &  1.6019  \\
7295.91   &  0.04$\pm$0.01   &  \ion{Mg}{2}$\lambda$2803.53    &  1.6024  \\
7296.85   &  0.23$\pm$0.01   &  \ion{Mg}{2}$\lambda$2803.53    &  1.6028  \\
7422.05   &  0.04$\pm$0.01   &  \ion{Mg}{1}$\lambda$2852.94    &  1.6014  \\
7423.01   &  0.23$\pm$0.01   &  \ion{Mg}{1}$\lambda$2852.94    &  1.6019  \\
\hline
\end{tabular}
\normalsize

$^a$ the equivalent width of unidentified lines is computed at
zero redshift.
? = uncertain identification. In many cases several components contribute to 
a single line in one entry of the table. In these cases either the redshift
is given with only 4 digits or it is given as a range. 
A + sign indicates that
a line is partly blended with the following entry.
\end{table}

\setcounter{table}{2}

\begin{table}[ht!]
\caption{\bf GRB 021004 UVES line identifications, continued.}
\footnotesize
\begin{tabular}{lccc}
\hline
\hline
$\lambda$(\AA\ ) & $W_{rest}$(\AA\ )$^a$&   ID  &    z  \\
\hline
7929.74   &  0.75$\pm$0.05   &                    &          \\
7929.53   &  0.08$\pm$0.02   &  \ion{Fe}{2}$\lambda$2382.77    &  2.328   \\
7930.42   &  0.04$\pm$0.02   &  \ion{Fe}{2}$\lambda$2382.77    &  2.328   \\
7931.03   &  0.03$\pm$0.02   &  \ion{Fe}{2}$\lambda$2382.77    &  2.328   \\
8228.16   &  0.37$\pm$0.05   &                    &          \\
9178.45   &  0.75$\pm$0.08   &                    &          \\
9205.45   &  0.70$\pm$0.09   &                    &          \\
9216.41   &  0.04$\pm$0.03   &  \ion{Mg}{2}$\lambda$2796.35    &  2.2958  \\
9222.70   &  0.09$\pm$0.03   &  \ion{Mg}{2}$\lambda$2796.35    &  2.2981  \\
9223.62   &  0.18$\pm$0.03   &  \ion{Mg}{2}$\lambda$2796.35    &  2.2984  \\
9246.33   &  0.09$\pm$0.03   &  \ion{Mg}{2}$\lambda$2803.53    &  2.2981  \\
9247.39   &  0.12$\pm$0.03   &  \ion{Mg}{2}$\lambda$2803.53    &  2.2984  \\
9330.33   &  0.57$\pm$0.03   &  \ion{Mg}{2}$\lambda$2803.53    &  2.328   \\
\hline
\end{tabular}
\normalsize

$^a$ the equivalent width of unidentified lines is computed at
zero redshift.
\end{table}

\begin{table}[ht!]
\caption{\bf GRB 020813 UVES line identifications}
\footnotesize
\begin{tabular}{lccc}
\hline
\hline
$\lambda$(\AA\ ) & $W_{rest}$(\AA\ )$^a$&   ID  &    z  \\
\hline
4824.64  &  1.06$\pm$     0.05  &                                  &      \\
4824.64  &  0.59$\pm$     0.05  &                                  &      \\
4998.03  &  0.47$\pm$     0.05  &   	    & \\
5212.23	 &  0.18$\pm$     0.03  &   \ion{Fe}{2}$\lambda$2344.21  & 1.2234 \\
5236.74  &  1.64$\pm$     0.07  &                                &        \\
5281.33  &  0.09$\pm$     0.03  &   \ion{Fe}{2}$\lambda$2344.21  & 1.2529 \\
5285.19  &  1.70$\pm$     0.03  &   \ion{Fe}{2}$\lambda$2344.21  & 1.255  \\
5297.11  &  0.85$\pm$     0.03  &   \ion{Fe}{2}$\lambda$2382.77  & 1.2234 \\
5353.61  &  1.31$\pm$     0.03  &   \ion{Fe}{2}$\lambda$2374.46  & 1.255  \\
5368.21  &  0.28$\pm$     0.03  &   \ion{Fe}{2}$\lambda$2382.77  & 1.2529 \\
5372.23  &  1.63$\pm$     0.03  &   \ion{Fe}{2}$\lambda$2382.77  & 1.255  \\
5840.73  &  0.44$\pm$     0.06  &   	    & \\
5850.86  &  0.73$\pm$     0.06  &   	    & \\
5858.24  &  0.23$\pm$     0.02  &   \ion{Fe}{2}$\lambda$2600.17  & 1.2529 \\
5862.33  &  1.82$\pm$     0.02  &   \ion{Fe}{2}$\lambda$2600.17  & 1.255  \\
5896.25  &  0.40$\pm$     0.06  &   	    & \\
6067.05  &  2.30$\pm$     0.06  &                                &        \\
6216.93  &  1.34$\pm$     0.03  &   \ion{Mg}{2}$\lambda$2796.35  & 1.2234 \\ 
6232.85  &  1.36$\pm$     0.03  &   \ion{Mg}{2}$\lambda$2803.53  & 1.2234 \\ 
6300.13  &  0.55$\pm$     0.03  &   \ion{Mg}{2}$\lambda$2796.35  & 1.2529 \\ 
6304.72  &  2.38$\pm$     0.03  &   \ion{Mg}{2}$\lambda$2796.35  & 1.255  \\ 
6310.38  &  0.30$\pm$     0.06  &                                &        \\
6316.13  &  0.30$\pm$     0.03  &   \ion{Mg}{2}$\lambda$2803.53  & 1.2529 \\ 
6321.11  &  2.19$\pm$     0.03  &   \ion{Mg}{2}$\lambda$2803.53  & 1.255  \\ 
6343.49  &  0.11$\pm$     0.03  &   \ion{Mg}{1}$\lambda$2852.13  & 1.2234 \\
6427.36  &  0.24$\pm$     0.03  &   \ion{Mg}{1}$\lambda$2852.13  & 1.2529 \\ 
6432.12  &  1.34$\pm$     0.03  &   \ion{Mg}{1}$\lambda$2852.13  & 1.255  \\ 
6686.87  &  0.27$\pm$     0.03  &   \ion{Fe}{1}$\lambda$2967.76  & 1.2529 \\
\hline
\end{tabular}
\normalsize

Errors are 67\% confidence intervals.
$^a$ the equivalent width of unidentified lines is computed at
zero redshift.
\end{table}

\begin{figure*}
\centerline{ 
\vbox{
\psfig{figure=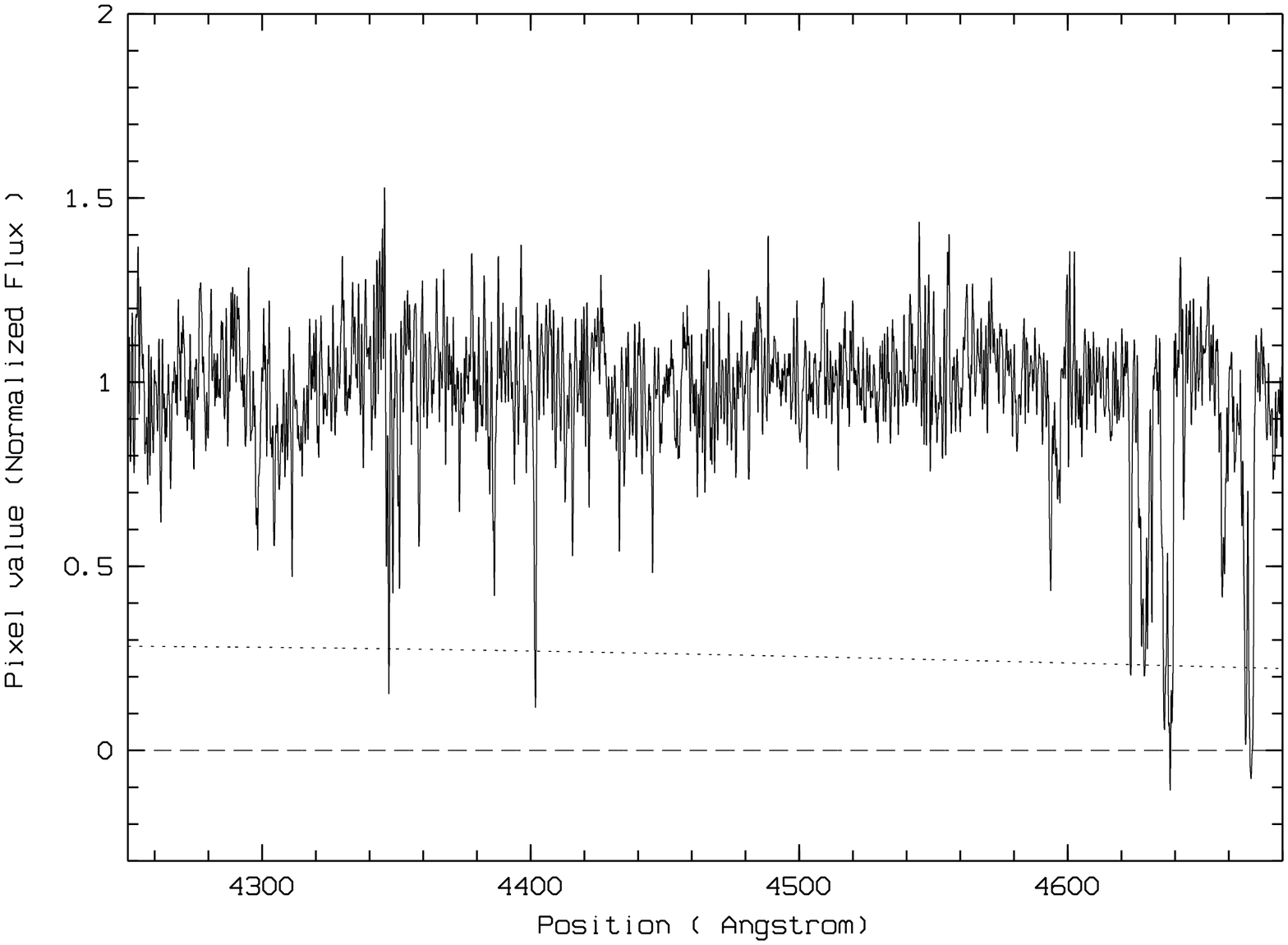,height=16cm,width=5.5cm,angle=-90}
\psfig{figure=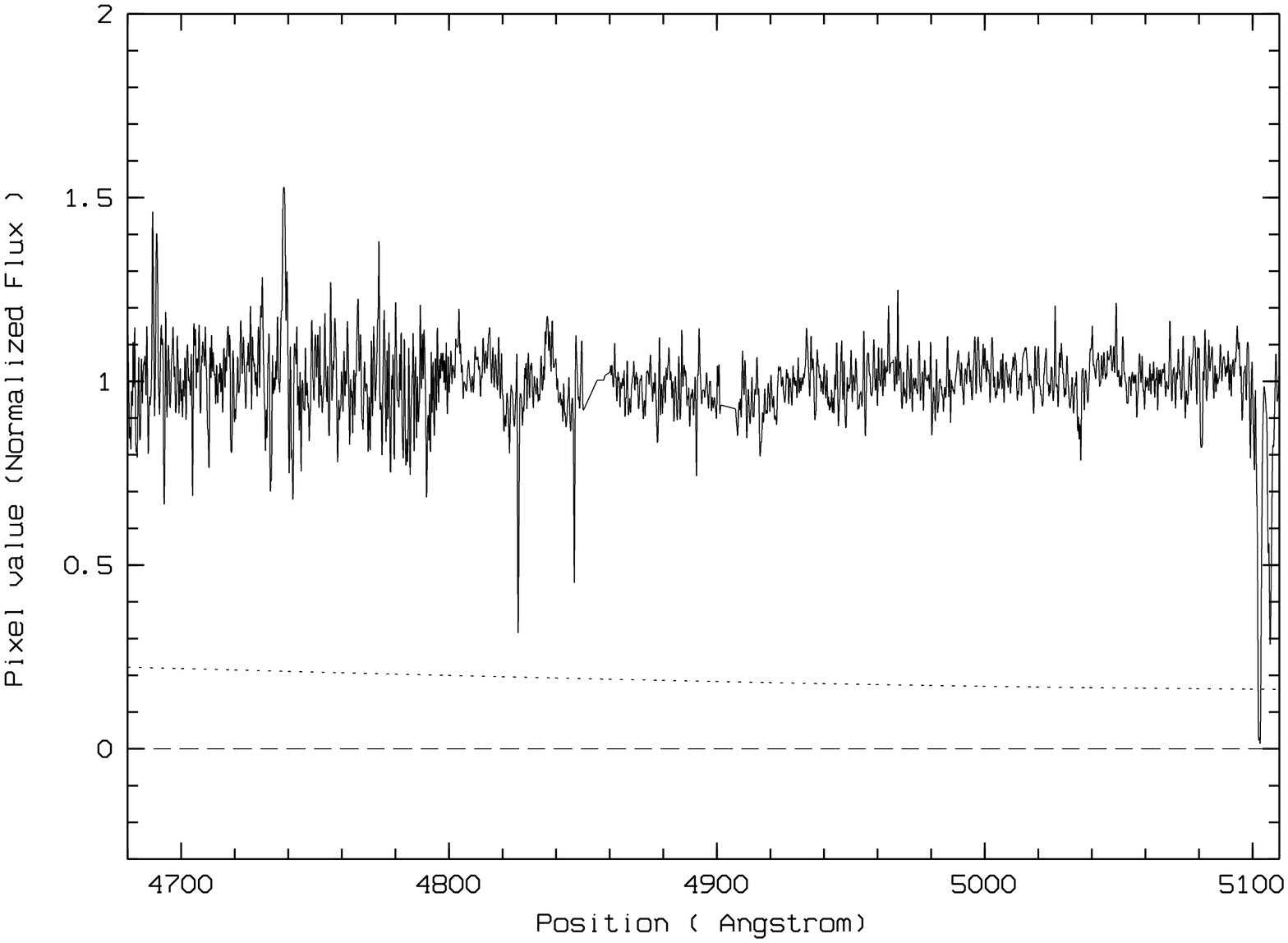,height=16cm,width=5.5cm,angle=-90}
\psfig{figure=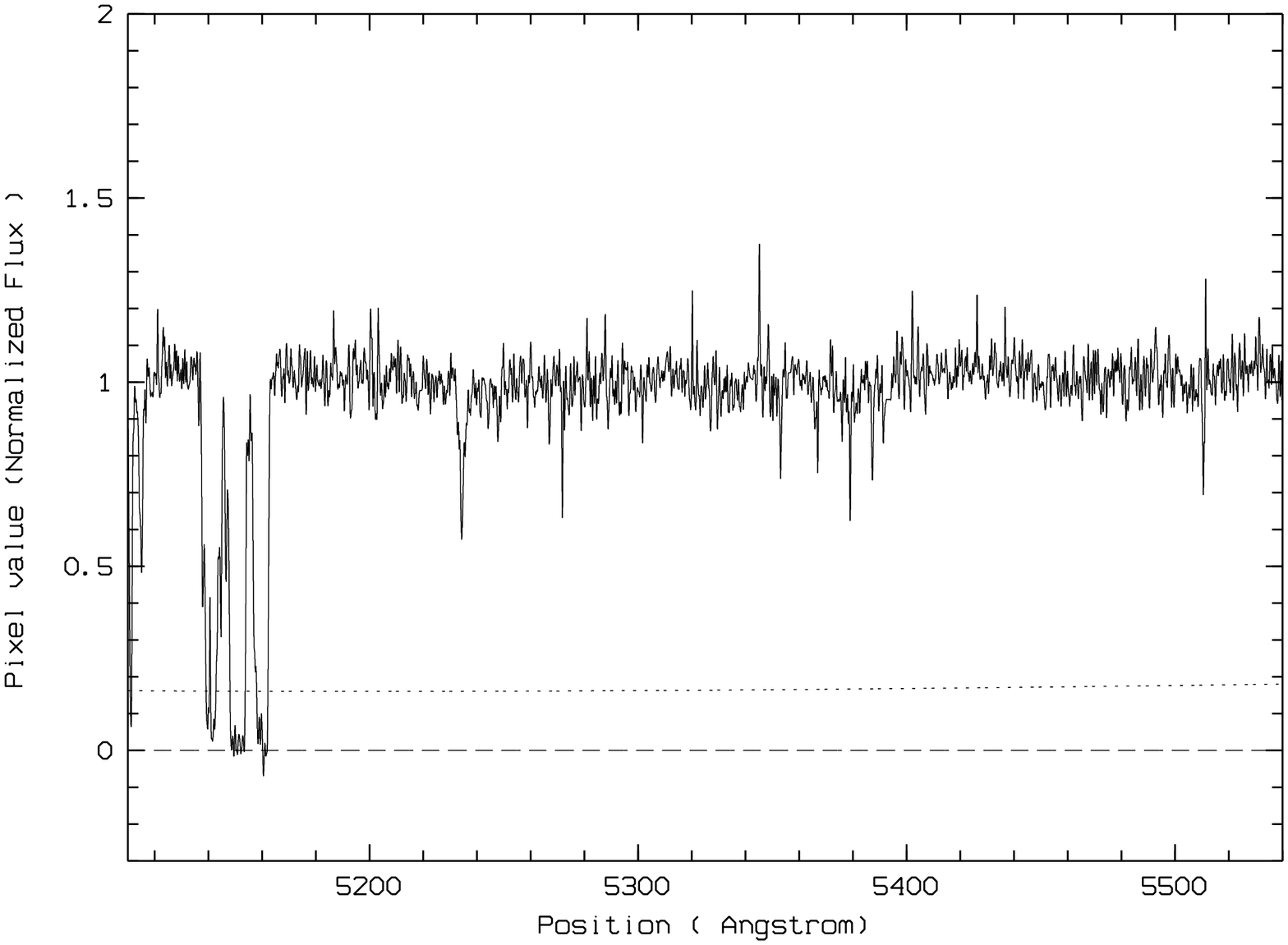,height=16cm,width=5.5cm,angle=-90}
\psfig{figure=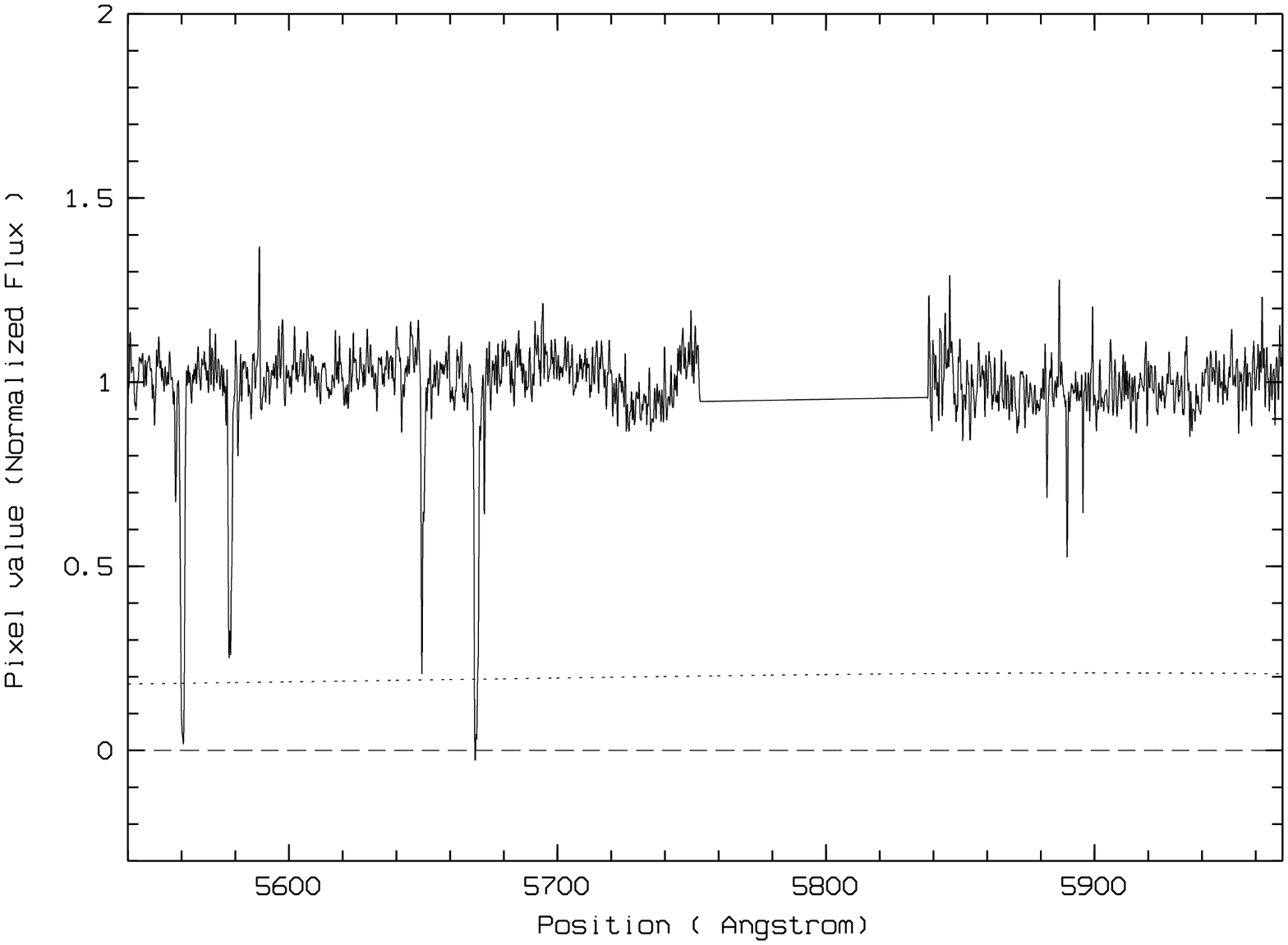,height=16cm,width=5.5cm,angle=-90}
}
}
\label{spe021004}
\caption {UVES spectrum of the GRB 021004 afterglow smoothed with a
gaussian function with $\sigma=1.5$ pixel . The dotted line is the 
error spectrum}
\end{figure*}

\setcounter{figure}{0}

\begin{figure*}
\centerline{ 
\vbox{
\psfig{figure=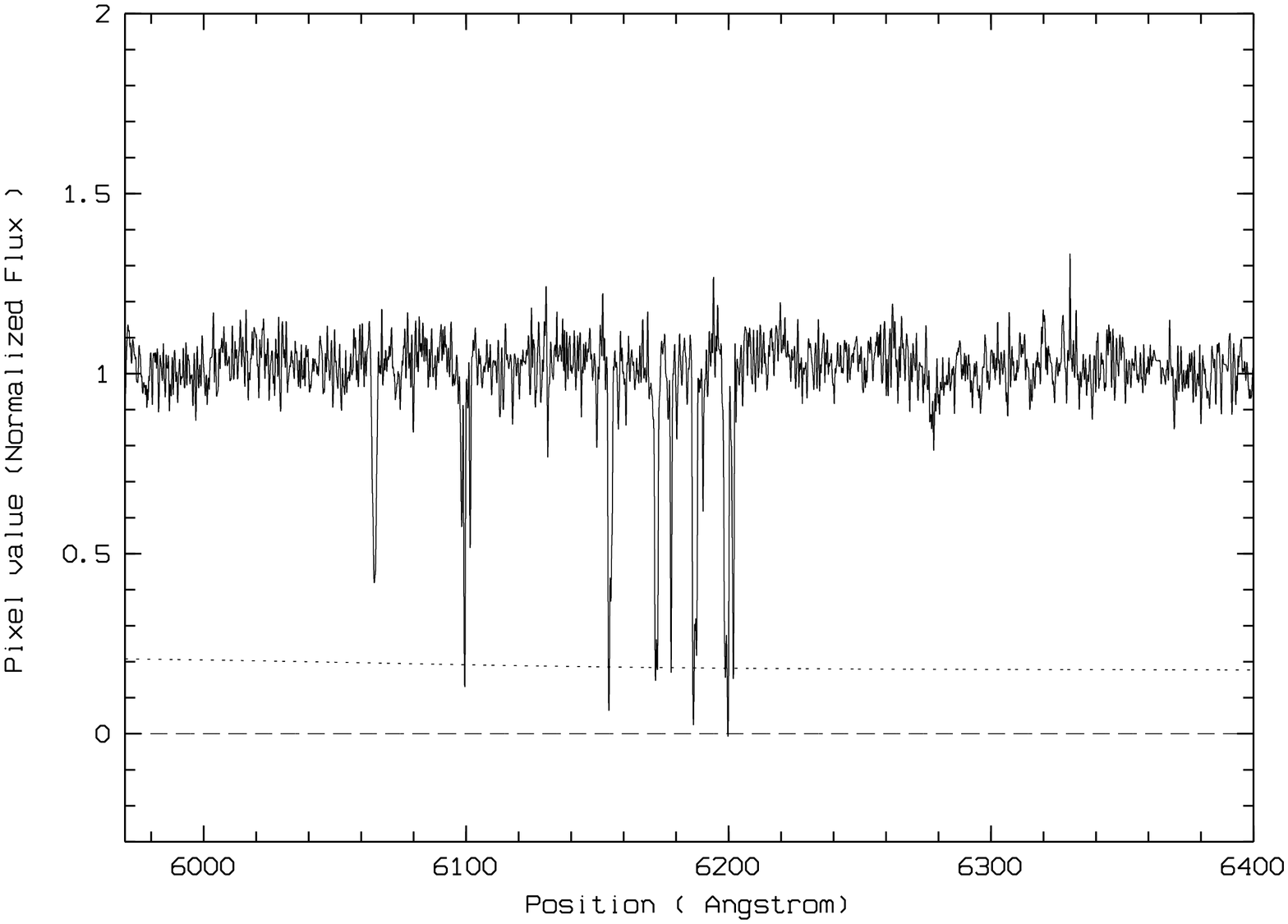,height=16cm,width=5.5cm,angle=-90}
\psfig{figure=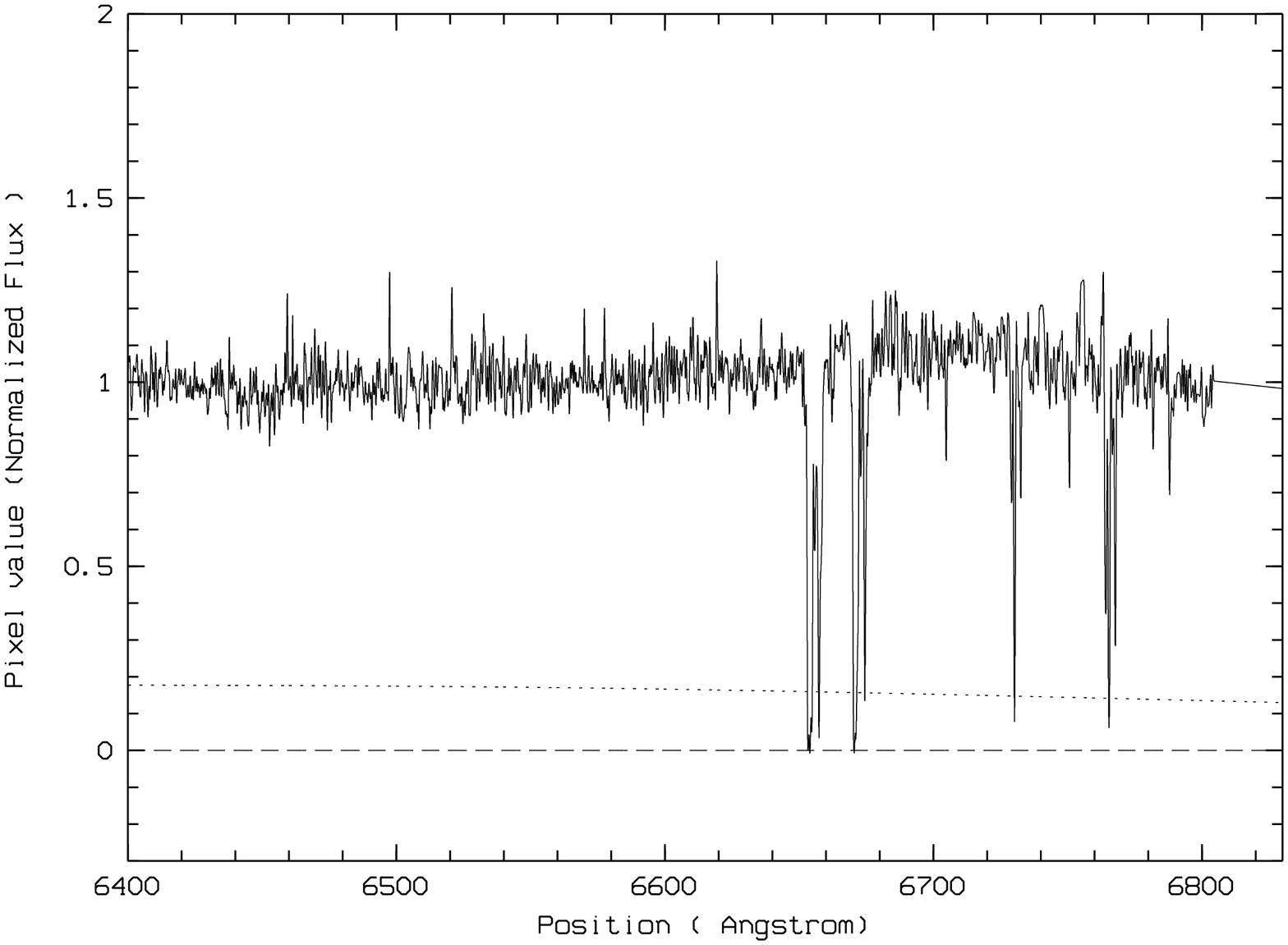,height=16cm,width=5.5cm,angle=-90}
\psfig{figure=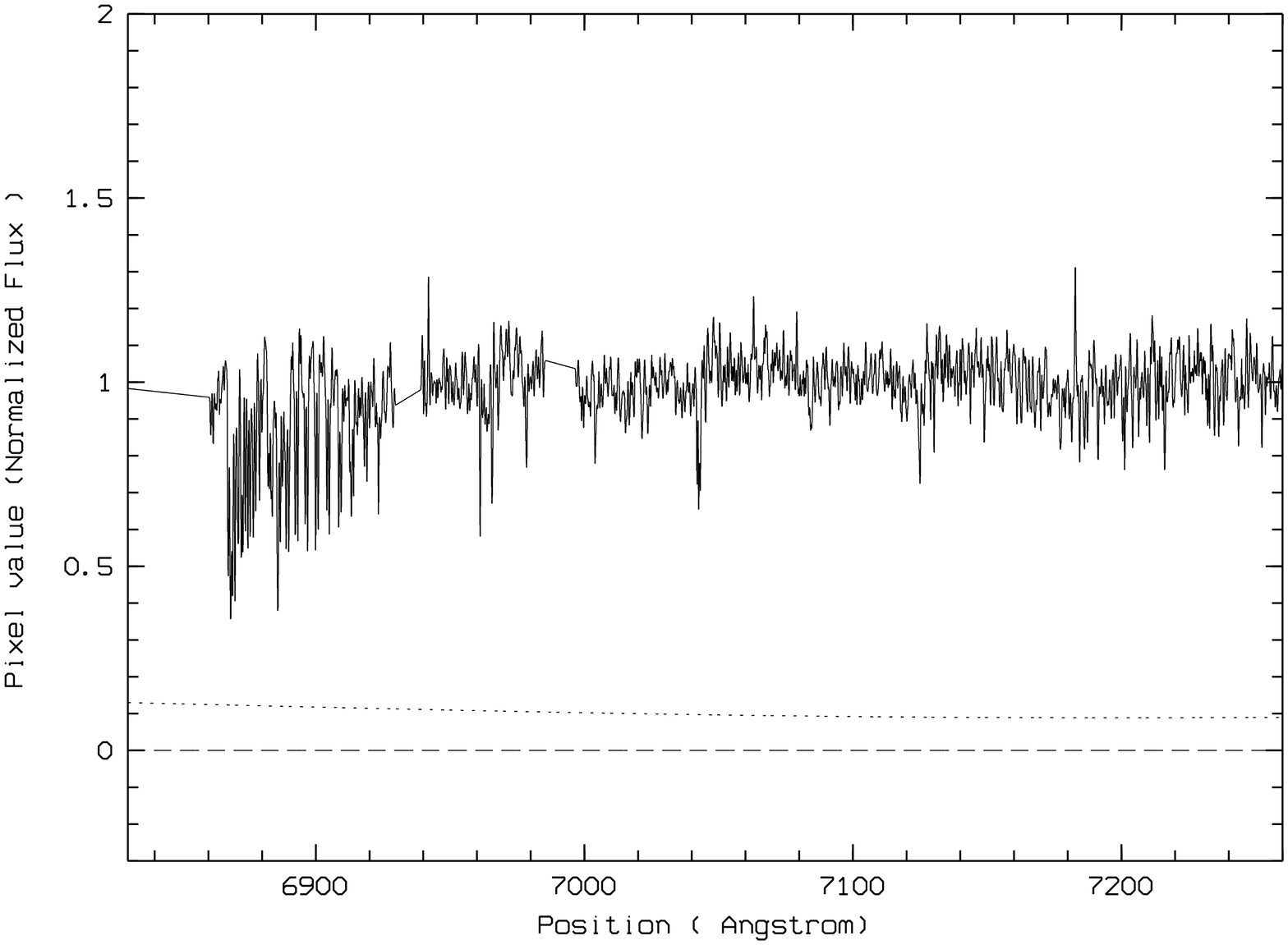,height=16cm,width=5.5cm,angle=-90}
\psfig{figure=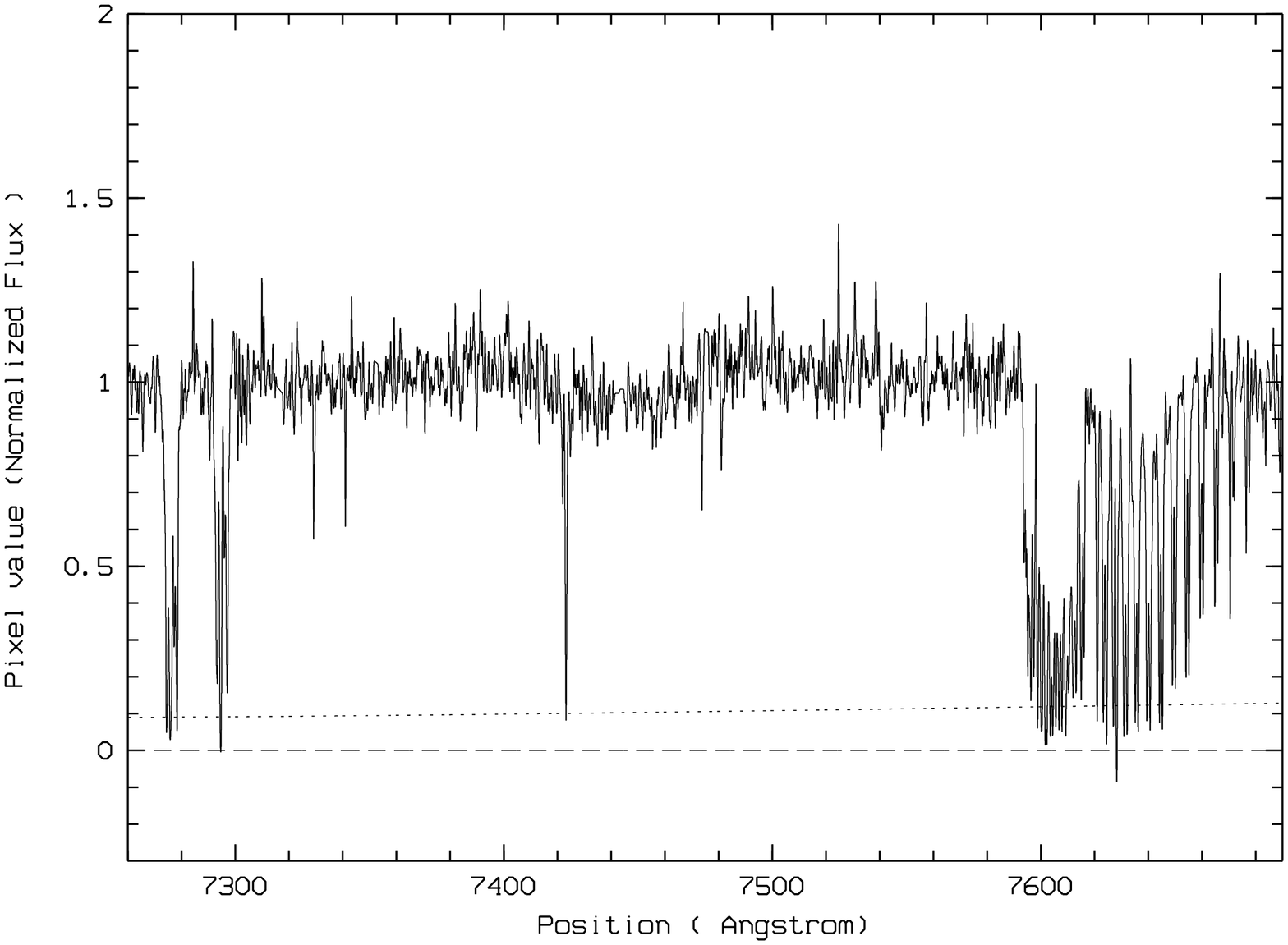,height=16cm,width=5.5cm,angle=-90}
}
}
\caption {continued}
\end{figure*}

\setcounter{figure}{0}

\begin{figure*}
\centerline{ 
\vbox{
\psfig{figure=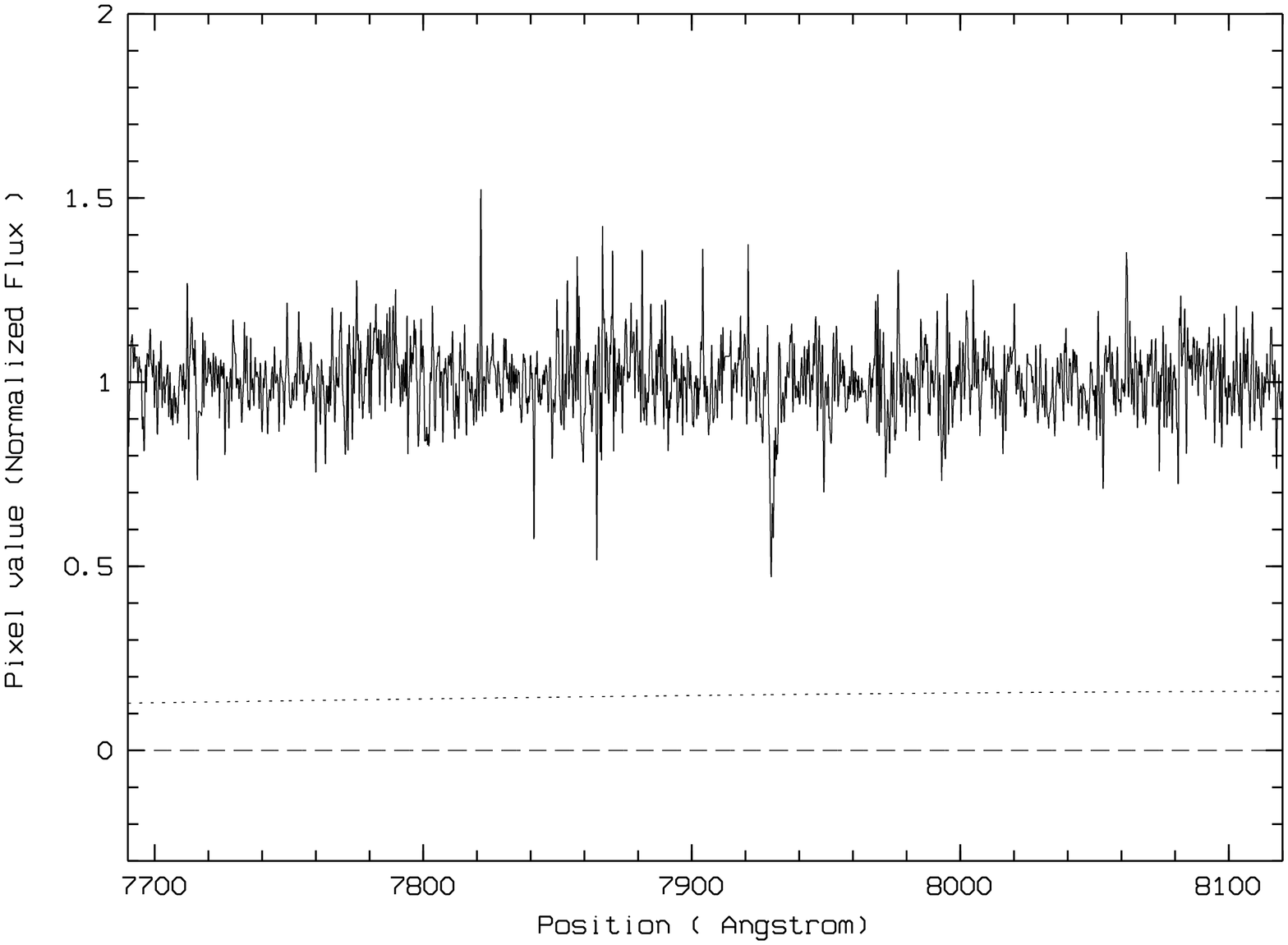,height=16cm,width=5.5cm,angle=-90}
\psfig{figure=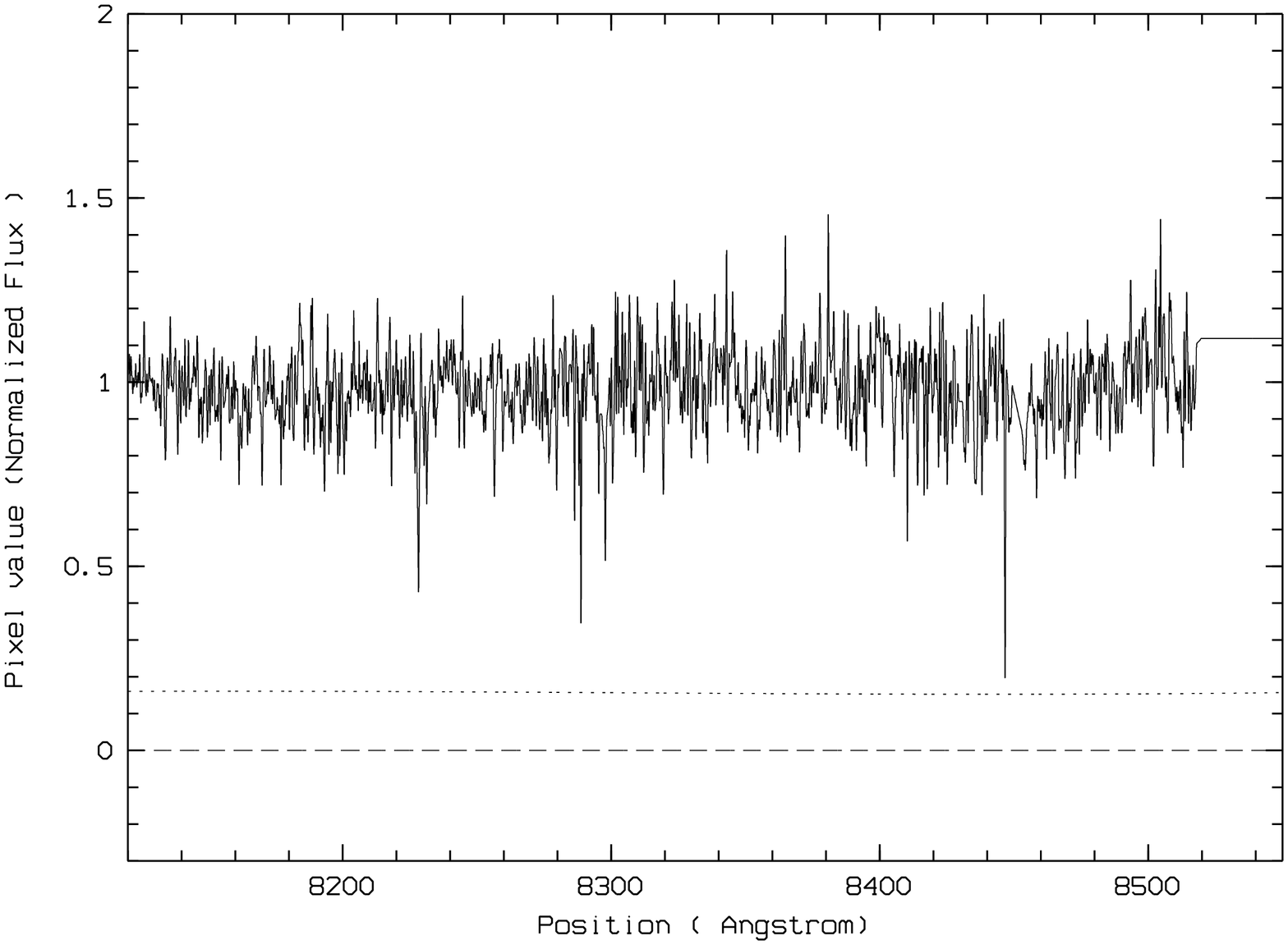,height=16cm,width=5.5cm,angle=-90}
\psfig{figure=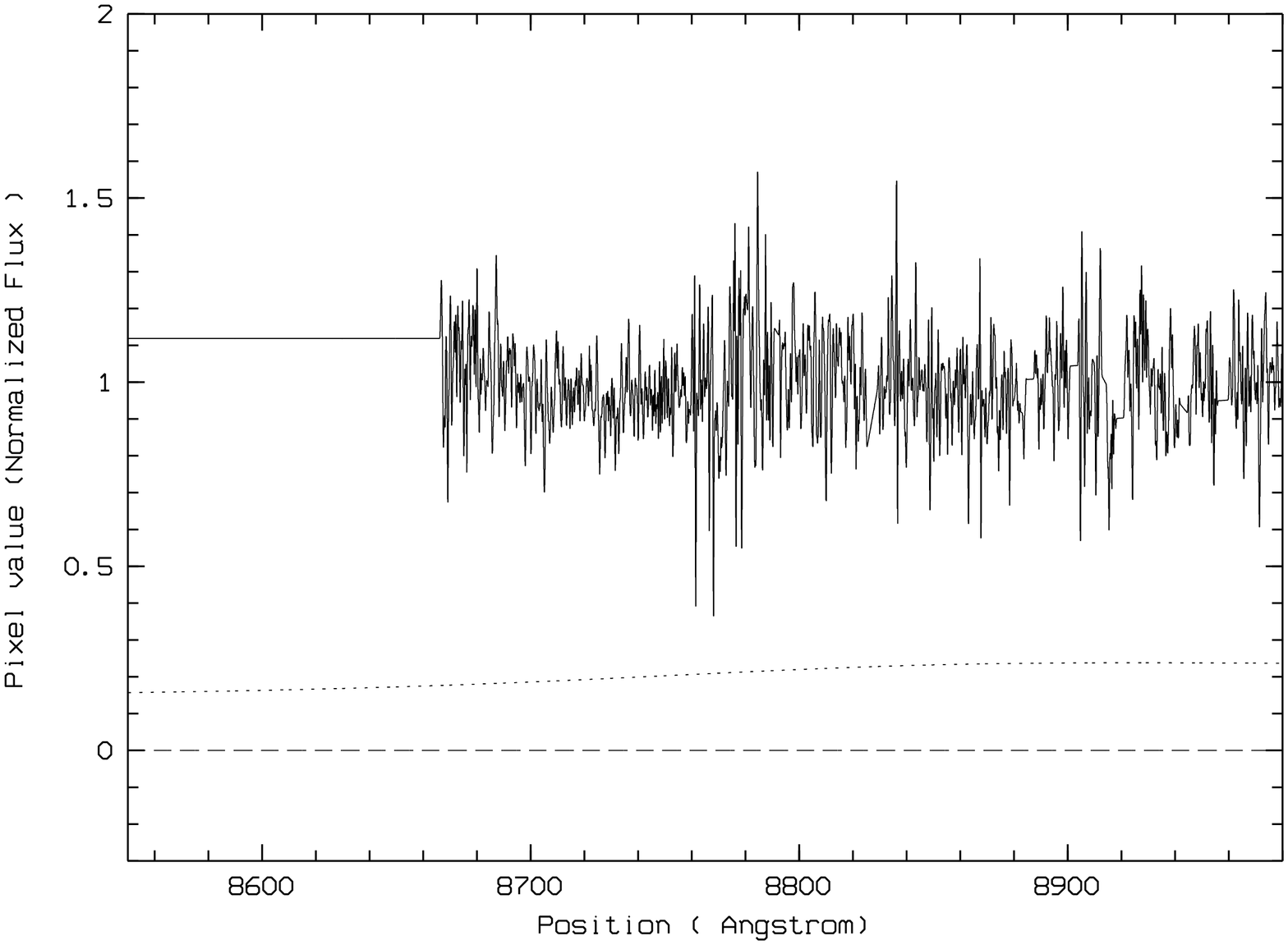,height=16cm,width=5.5cm,angle=-90}
\psfig{figure=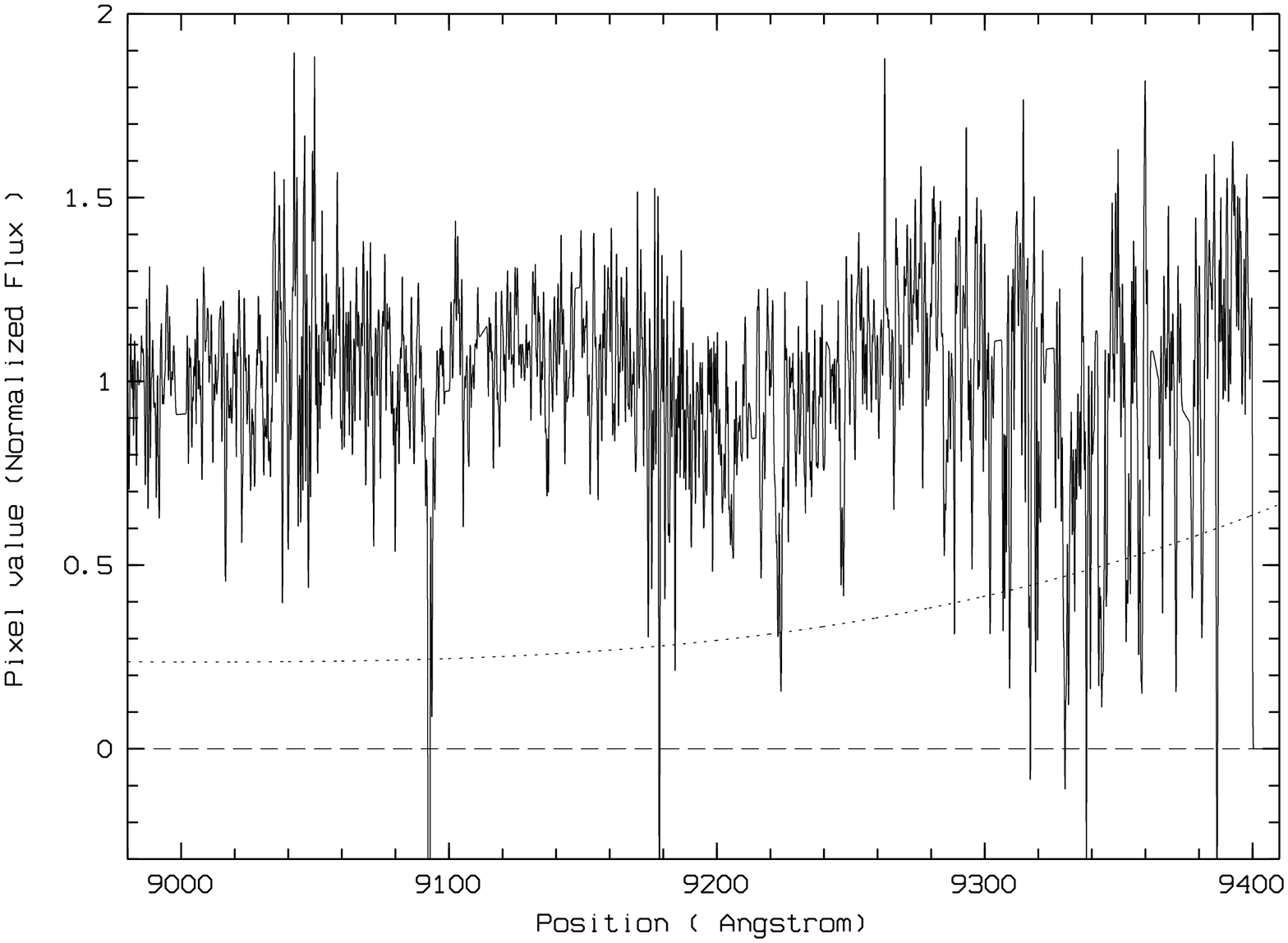,height=16cm,width=5.5cm,angle=-90}
}
}
\caption {continued}
\end{figure*}

\begin{figure*}
\centerline{ 
\vbox{
\psfig{figure=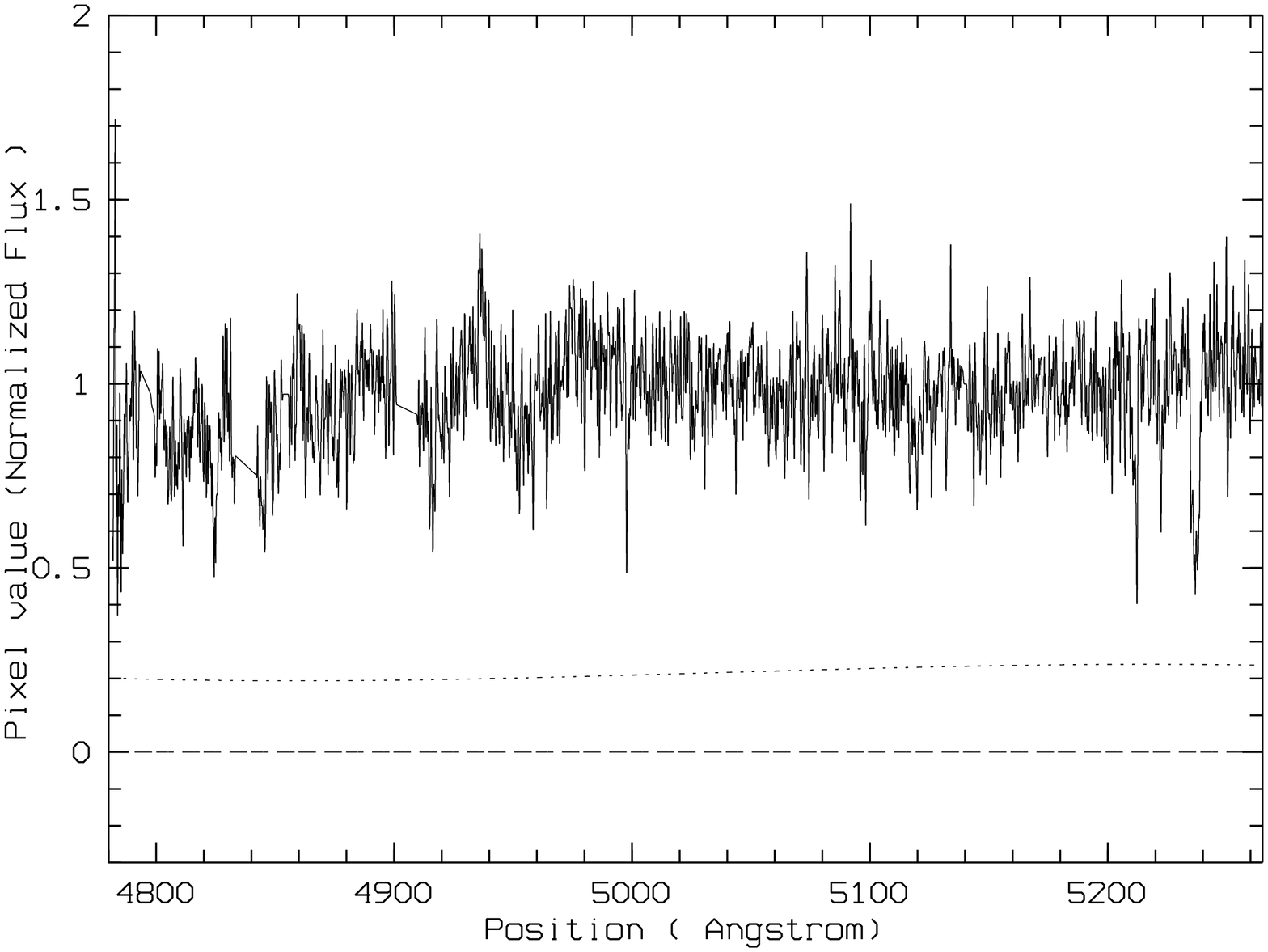,height=16cm,width=5.5cm,angle=-90}
\psfig{figure=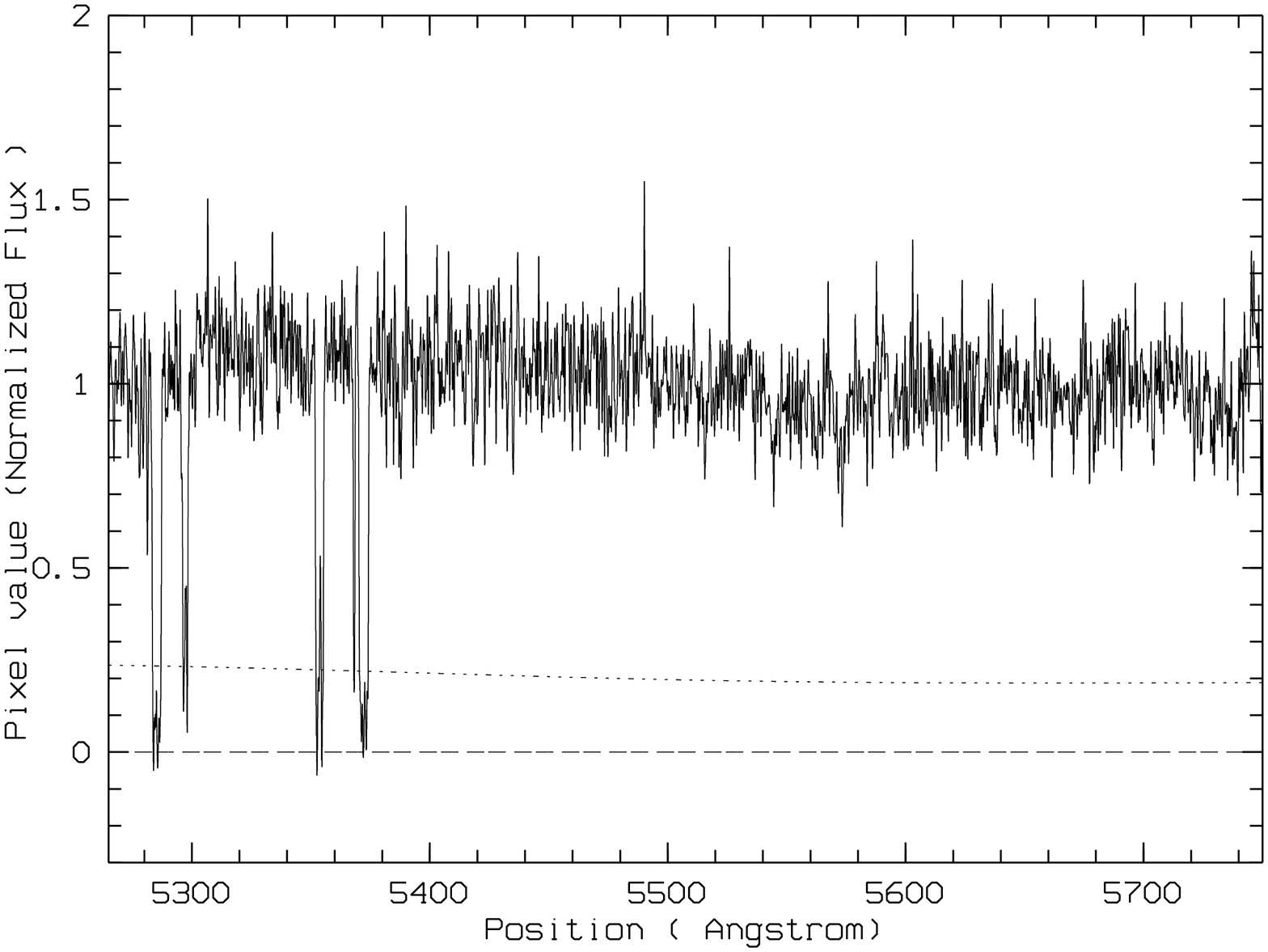,height=16cm,width=5.5cm,angle=-90}
\psfig{figure=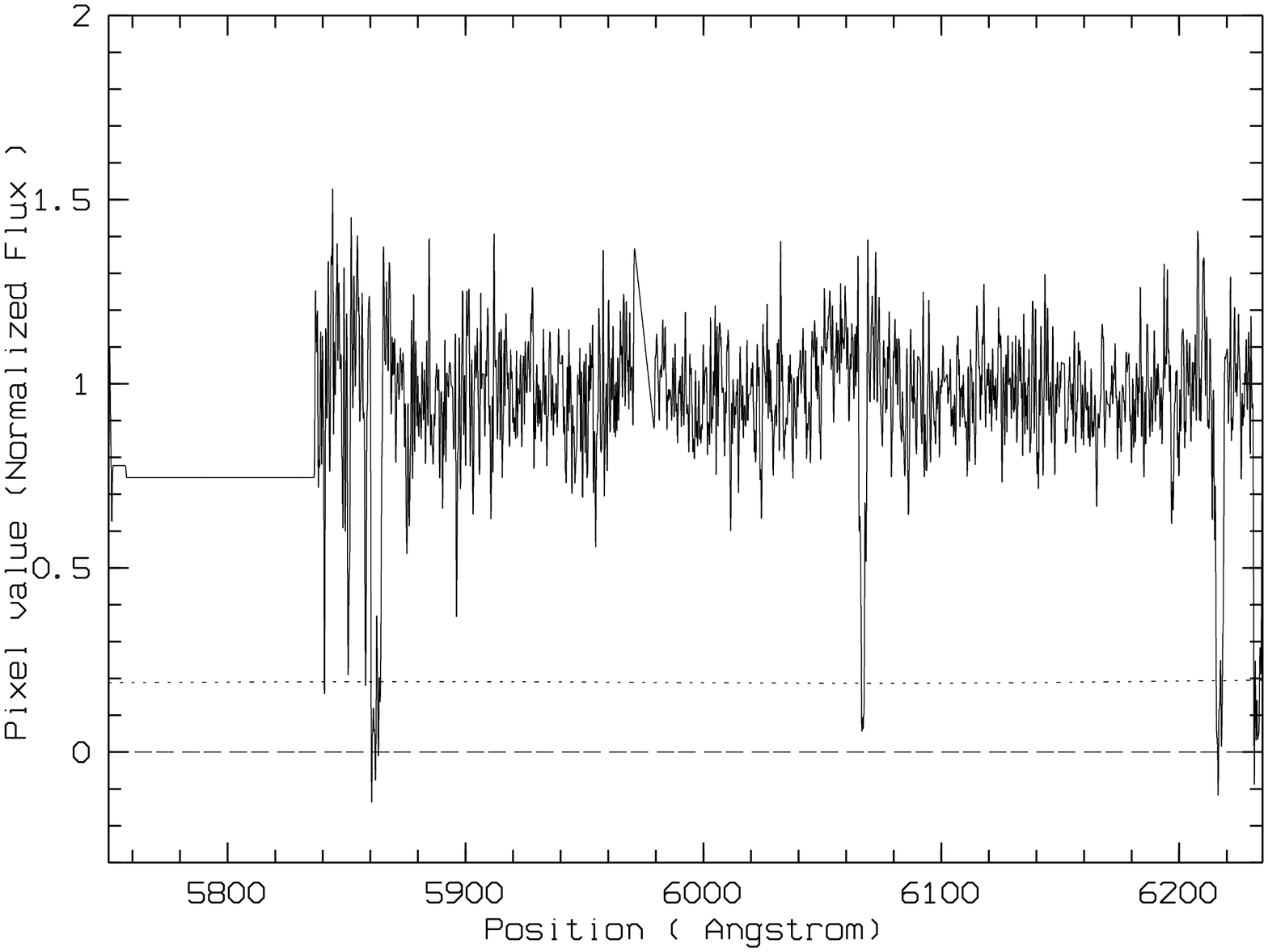,height=16cm,width=5.5cm,angle=-90}
\psfig{figure=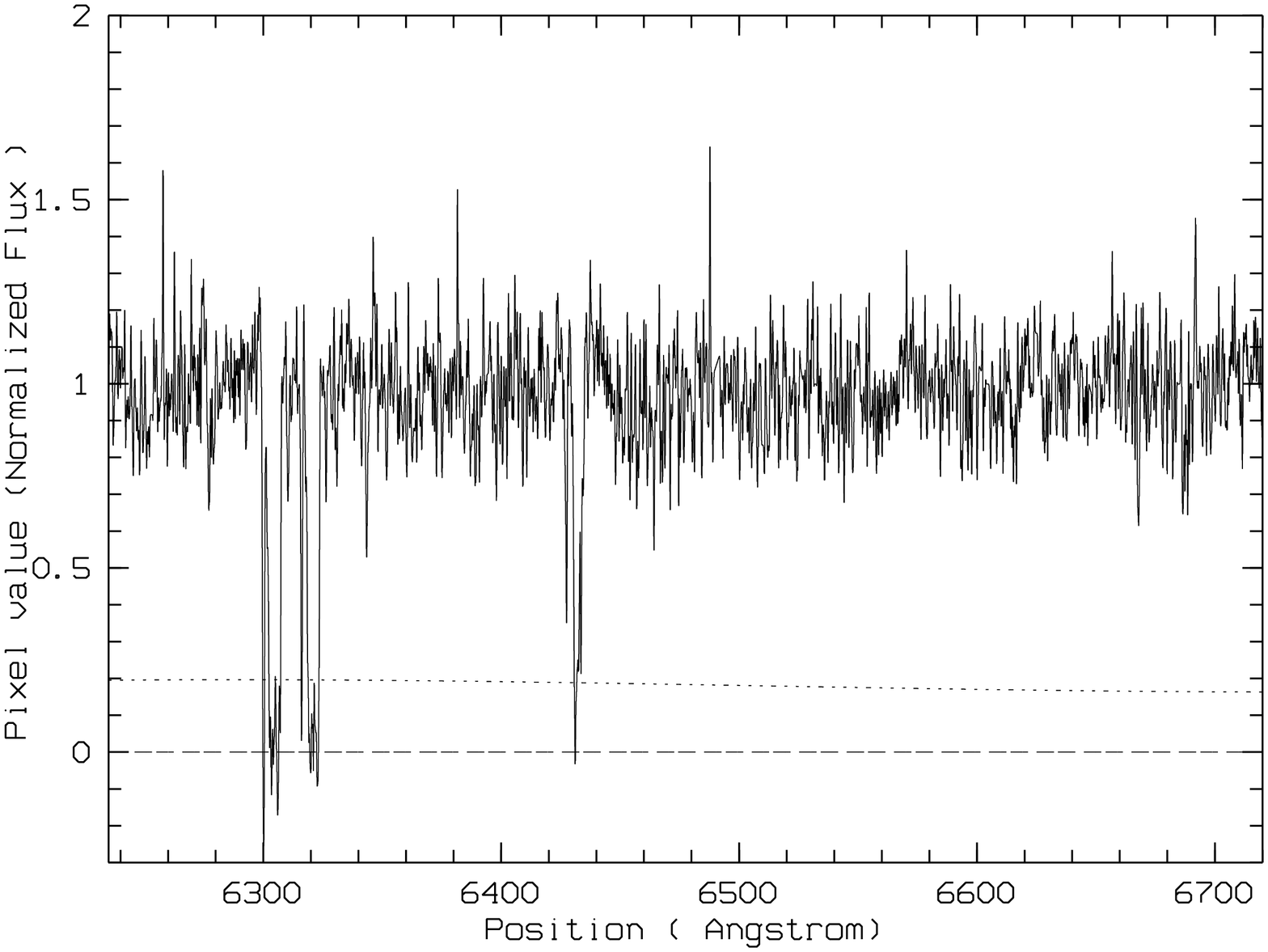,height=16cm,width=5.5cm,angle=-90}
}
}
\label{spe020813}
\caption {UVES spectrum of the GRB 020813 afterglow smoothed with a
gaussian function with $\sigma=1.5$ pixel . The dotted line is the 
error spectrum}
\end{figure*}

\section{Column densities}

In this paper we focus on the systems which are likely to be
associated with the GRB host galaxies. The line fitting was 
performed using the MIDAS package FITLYMAN (Fontana \& Ballester
1995). This uses a Voigt profile and yields independently
the column density N and the Doppler parameter $b$ for each absorption
component.  For each absorption system several lines, spread over the
entire spectral range covered by the UVES observations, were 
fitted simultaneously, using the same number of components for each
line, and the same redshift and $b$ value for each component.

\subsection{GRB 021004}

\begin{figure*}
\centerline{ 
\vbox{
\psfig{figure=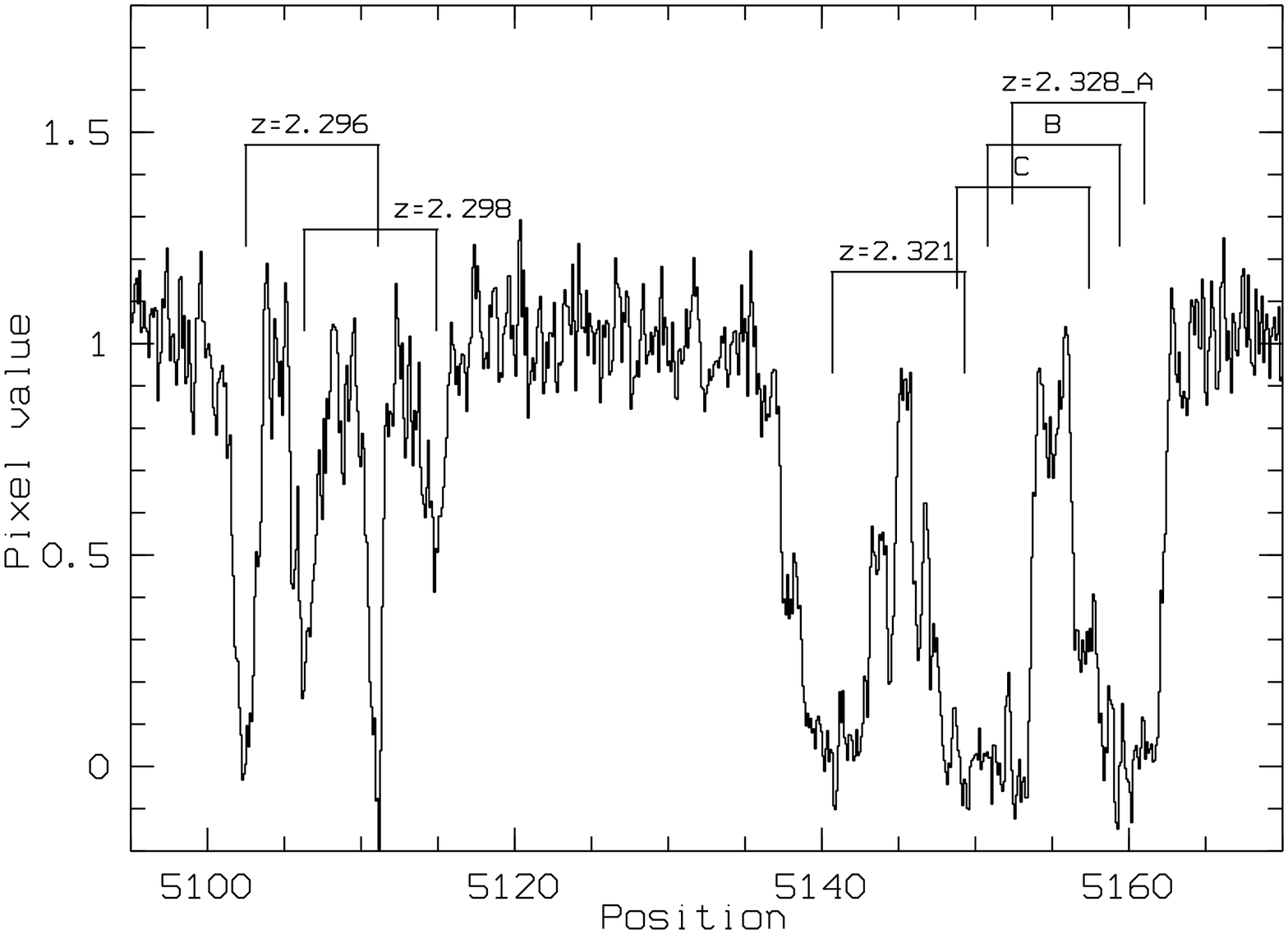,height=14cm,width=10cm,angle=-90}
\psfig{figure=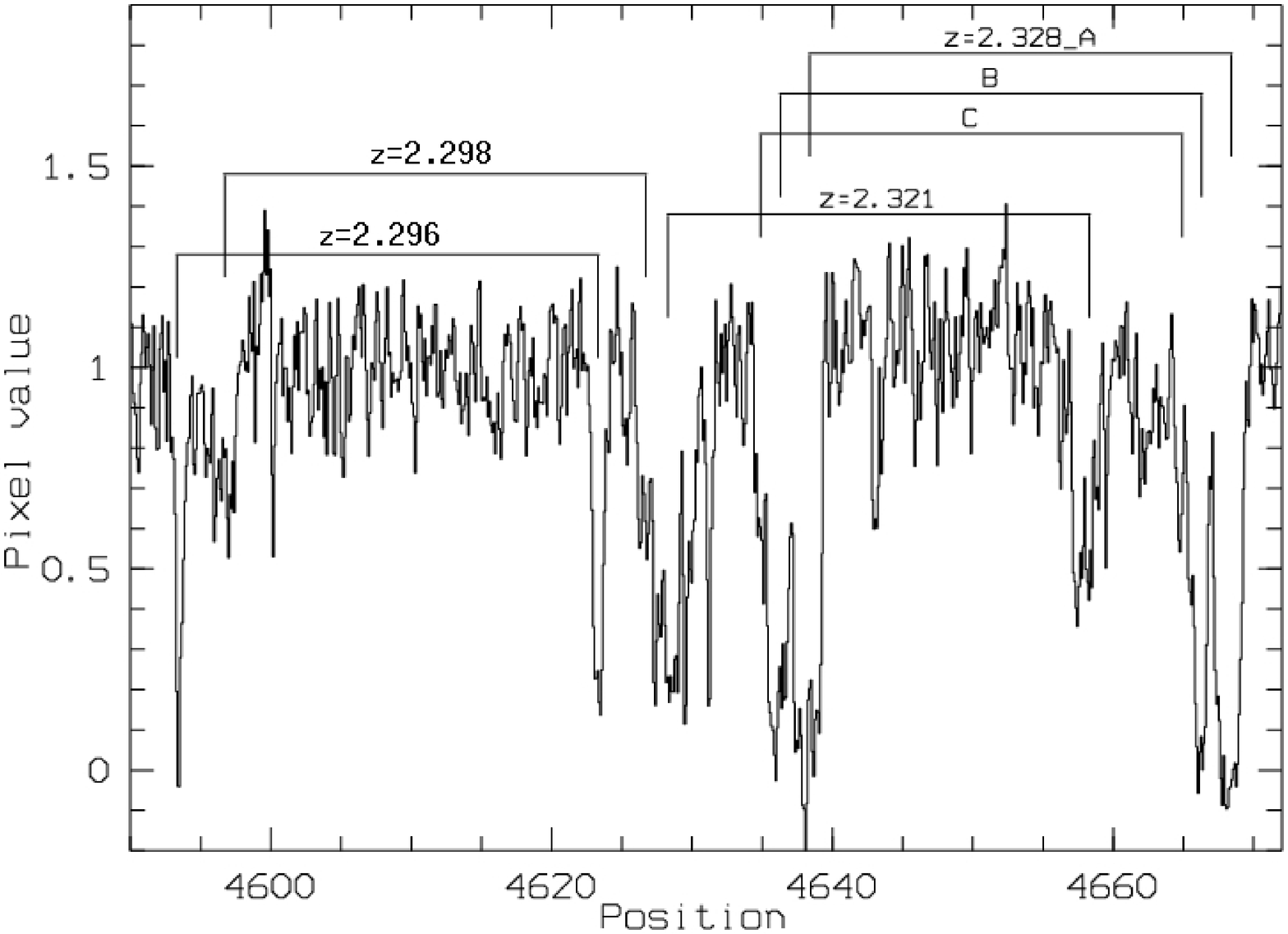,height=14cm,width=10cm,angle=-90}
}
}
\caption{\ion{C}{4} (upper panel) and \ion{Si}{4} (lower panel) absorption
systems in the UVES spectra of GRB 021004 for z=2.328 (see Table 5). 
}
\label{sys}
\end{figure*}

For this GRB we consider the absorption systems at the following
redshifts and velocities (in km/s) with respect to the Ly$\alpha$
emission of the host galaxy (Mirabal et al. 2003): z=2.328, v=0
(system A in figure \ref{sys}); z=2.328, v=--139 km/s (system B);
z=2.328, v=--224 km/s (system C); z=2.321; z=2.298; z=2.296.  For
these systems we detected \ion{C}{4}, \ion{C}{2}, \ion{Si}{4},
\ion{Al}{2}, \ion{Al}{3}, \ion{Mg}{2} and \ion{Fe}{2} lines (see Table
3).  We report in the table also the tentative identification of
a \ion{Si}{2}$\lambda$1304 line at z=2.2953. This is close but not
coincident with the redshift of systems z=2.296. We therefore consider
this identification uncertain, and we do not consider this line in
the following analysis.

Unfortunately, the Ly${\alpha}$
absorption associated with the z=2.328 systems falls exactly in the gap
between the dichroic 1 and 2 blue arm spectra, and therefore is not
covered in this study. \ion{Zn}{2} and \ion{Cr}{2} absorption lines
for the main z=2.328 and z=2.321 systems fall in the region affected
by atmospheric telluric features (see figure 1) and are
therefore not accessible.

Figure \ref{sys} shows the \ion{C}{4} and \ion{Si}{4} doublets of the
above 6 systems. Note that the \ion{C}{4}$\lambda$1548 lines of the
z=2.328 A,B,C systems are strongly blended with the
\ion{C}{4}$\lambda$1550 line of the z=2.321 system, while the
\ion{Si}{4}$\lambda$1404 line of the z=2.298 system is blended with
the \ion{Si}{4}$\lambda$1393 line of the z=2.321 system.  Each of the
six systems actually comprises several components, within a velocity
range of several tens of km/s. For this reason the identification of
the different systems is somewhat subjective, the true message being
that the geometry and kinematics of the ISM clouds probed by the GRB
line of sight are complex.  Nevertheless, sticking to the above system
identifications will be useful.

There are 13 lines in Table 3 associated with the z=2.328
systems, 5 lines associated with the z=2.321 system, and up to 17 lines
associated to the z=2.296-2.298 systems. Some of these lines can be
split further in several components. To test the robustness of the
fit, in terms of accuracy and stability of the results, we performed
many fits, using several combination of lines/systems. We found that
the fits presented below are a good compromise between increasing the
statistical precision of the fit, obtained increasing the number of
lines fitted simultaneously, and the stability/repeatibility of the
results, which degraded increasing the number of fitted parameters, due
to the increasingly complex shape of the $\chi^2$ hyper-surface in the
parameter space, that may contain many local minima.

\begin{figure*}
\centerline{ 
\vbox{
\psfig{figure=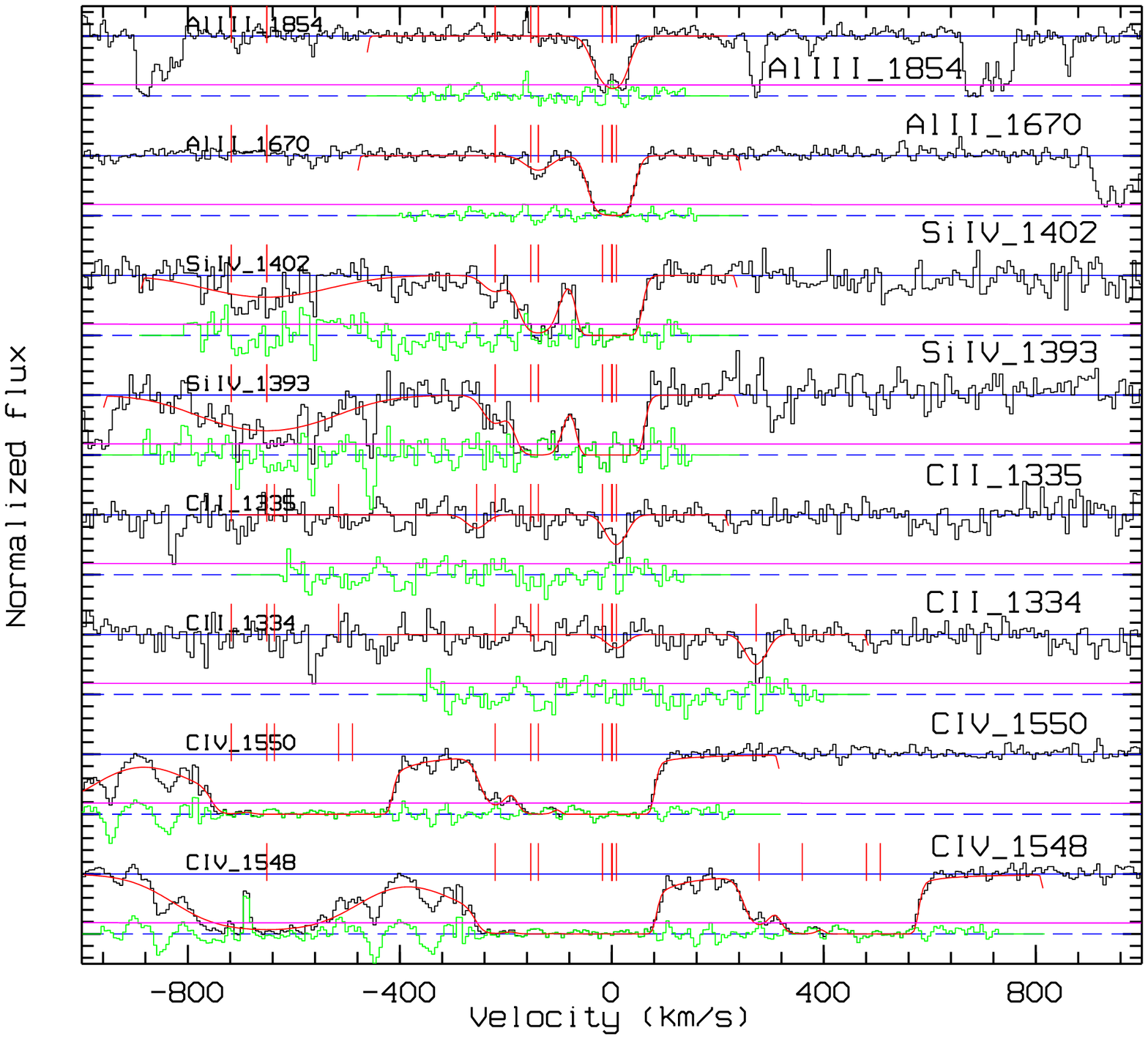,width=10cm,angle=-90}
\psfig{figure=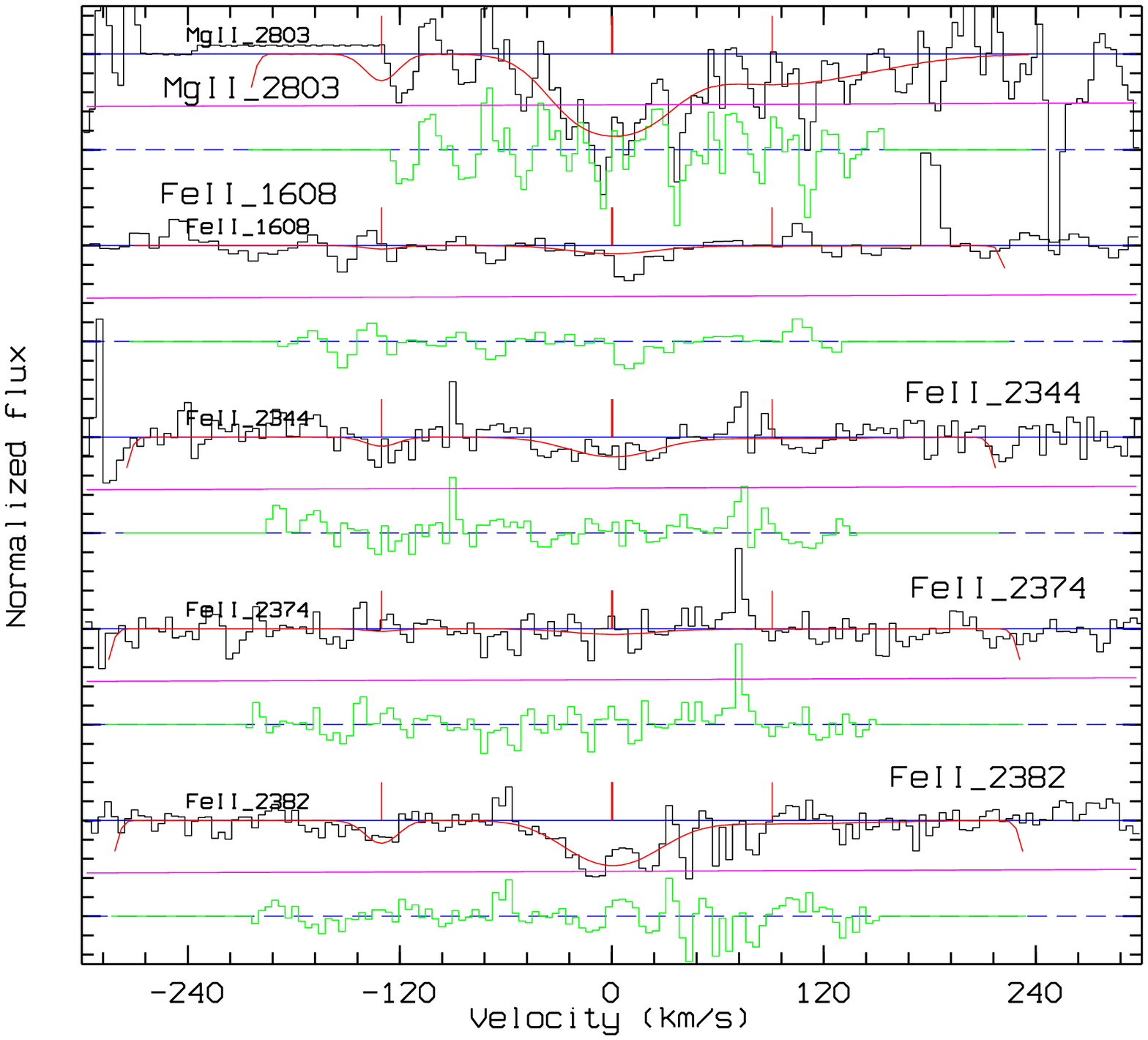,width=10cm,angle=-90}
}
}
\caption{GRB 021004. Upper panel: UVES spectrum near the \ion{C}{4}$\lambda
\lambda$1550, 1548, \ion{C}{2}$\lambda$1334, \ion{C}{2}$^*\lambda$1335,
\ion{Si}{4}$\lambda \lambda$1404, 1393,
\ion{Al}{2}$\lambda$1670 and \ion{Al}{3}$\lambda$1854
lines for the z=2.328 and z=2.321 systems in velocity space, along
with the best fit model (solid line) and residuals (grey). Lower
panel: same for the \ion{Fe}{2}$\lambda$2382,
\ion{Fe}{2}$\lambda$2374, \ion{Fe}{2}$\lambda$2344,
\ion{Fe}{2}$\lambda$1608 and \ion{Mg}{2}$\lambda$2803 lines.  The zero
in the velocity scale refers to the redshift of the host
galaxy.  }
\label{z1}
\end{figure*}

For the z=2.328 A,B,C and z=2.321 systems we fitted simultaneously the
\ion{C}{4}$\lambda \lambda$1550, 1548, \ion{C}{2}$\lambda$1334, 
\ion{C}{2}$^*\lambda$1335, \ion{Si}{4}$\lambda
\lambda$1404, 1393 \ion{Al}{2} $\lambda1670$ and the
\ion{Al}{3}$\lambda$1854 line. We excluded from the fit the z=2.328
\ion{Al}{3}$\lambda$1862 line because its blue wing is strongly
blended with another line, and the z=2.321 \ion{Al}{3}
$\lambda$1862 line because it is strongly blended with the
\ion{Fe}{2}$\lambda$2600 line of one of the components of the
z=1.38 intervening system.  For the same systems we also fitted the
\ion{Fe}{2}$\lambda$2382, \ion{Fe}{2}$\lambda$2374
\ion{Fe}{2}$\lambda$2344 \ion{Fe}{2}$\lambda$1608,
\ion{Mg}{2}$\lambda$2803 and \ion{Mg}{2}$\lambda$2796 lines
simultaneously. Figure \ref{z1} shows the line spectra in velocity
space, along with the best fit model, while Table 5 presents the best
fit abundances along with the velocity shift of each system with
respect to the redshift of the host galaxy, assumed to be 2.328 for
GRB 021004 (Mirabal et al. 2003).  We used 2 components for the
z=2.328\_A system, and 1 component for the other 3 systems. The best
fit doppler parameter b ranges from a minimum of 12$\pm$6 km/s (one of
the components of the z=2.328\_A system) up to a maximum of 107$\pm$10
km/s (system z=2.321). Of course different b values are obtained
using an higher (or lower) number of components. Unfortunately the
signal to noise ratio of our spectrum is not good enough to
unambiguously identify the different components. On the other hand, we
verified that the total best fit column density of each system is
stable, within the statistical errors, changing the number of
components in each system.

\begin{figure*}
\centerline{ 
\vbox{
\psfig{figure=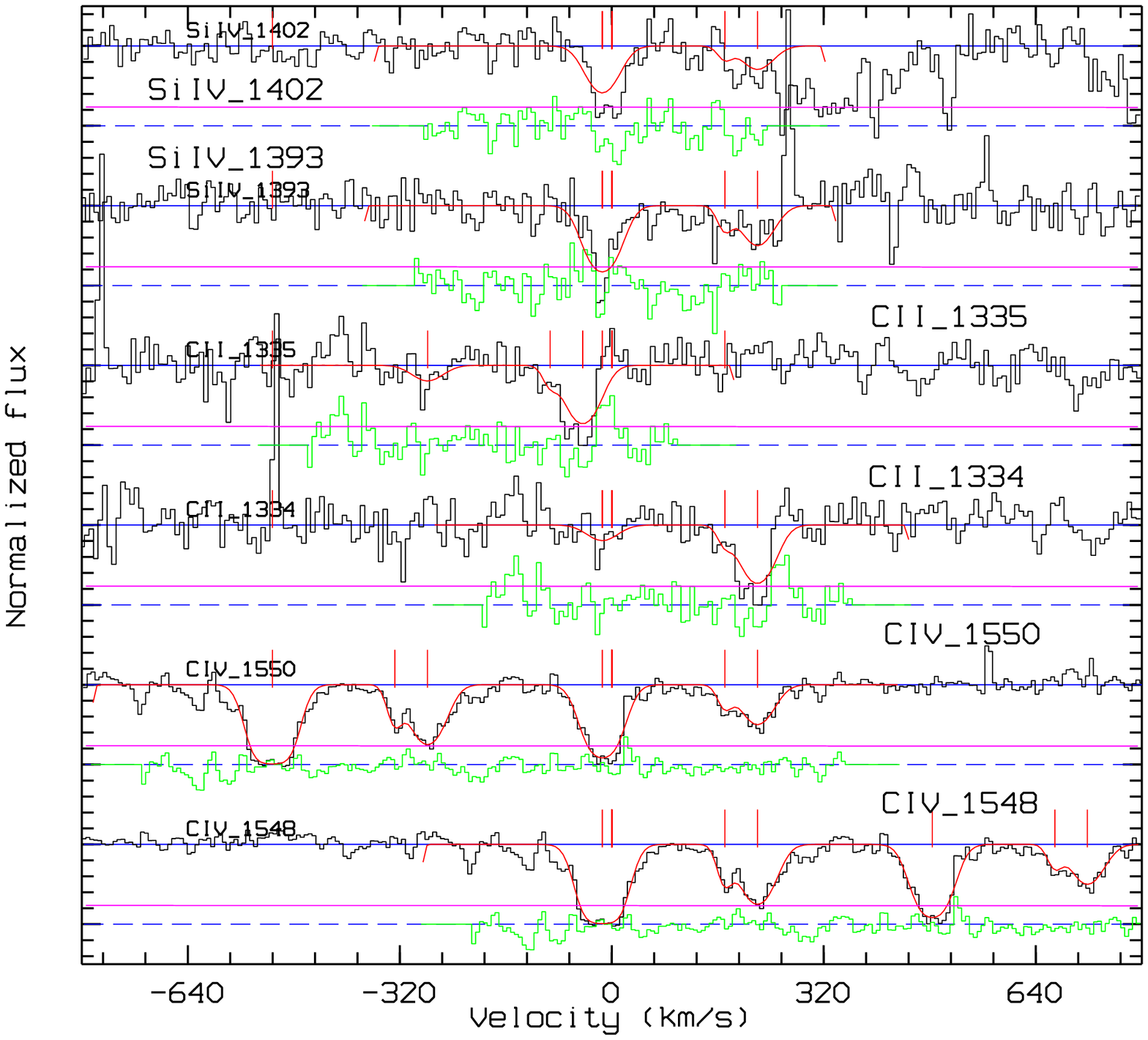,width=10cm,angle=-90}
\psfig{figure=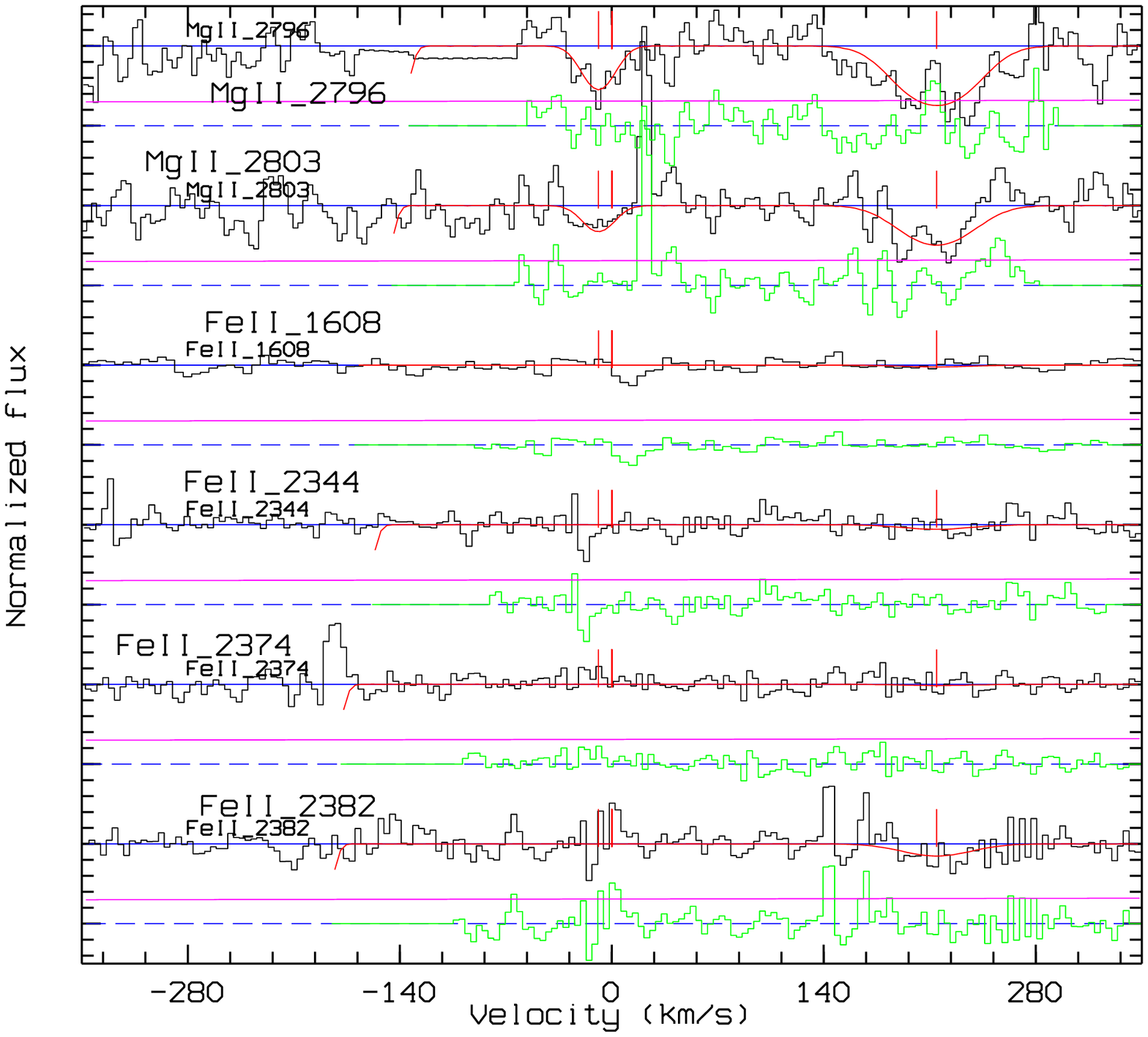,width=10cm,angle=-90}
}
}
\caption{GRB 021004. Upper panel: UVES spectrum near the \ion{C}{4} $\lambda
\lambda$ 1550, 1548, \ion{C}{2}$\lambda$1334, \ion{C}{2}$^*\lambda$1335,
\ion{Si}{4}$\lambda \lambda$1404, 1393
lines for the z=2.298 and z=2.296 systems in
velocity space, along with the best fit model (solid line) and
residuals (grey). Lower panel: same for the \ion{Fe}{2}$\lambda$2382,
\ion{Fe}{2}$\lambda$2374, \ion{Fe}{2}$\lambda$2344,
\ion{Fe}{2}$\lambda$1608, \ion{Mg}{2}$\lambda$2803 and
\ion{Mg}{2}$\lambda$2796 lines.  The zero in the velocity scale
refers to z=2.296.}
\label{z2}
\end{figure*}

Similar series of fits were performed for the z=2.296 and z=2.298
systems, for which we used 1 and 2 components, respectively (see
figure \ref{z2}).  Table 5 gives again the best fit abundances for
these 2 systems. The best fit doppler parameter b ranges from 
9$\pm$7 km/s to 40$\pm$13 km/s.

\begin{table}
\caption{\bf GRB 021004. Logarithmic ion column densities in cm$^{-2}$}
\footnotesize
\begin{tabular}{lcccccccc}
\tableline\tableline
System & v. shift$^{a}$ & \ion{Si}{4}  & \ion{C}{4}      & \ion{C}{2}      & \ion{C}{2}$^*$  & \ion{Al}{2}     & \ion{Fe}{2}     & \ion{Mg}{2} \\
\hline
2.328\_A & 0        & 15.30$\pm$0.56 &  $>$15.2        & 13.40$\pm$0.45  & 13.90$\pm$0.58 & 13.55$\pm$0.35$^c$ & 13.34$\pm$0.15 & 13.82$\pm$0.24\\
2.328\_B & --139    & 14.27$\pm$0.16 &  $>$14.4        & $<$13.2         & $<$13.2        & 12.23$\pm$0.35 & 12.53$\pm$0.24 & 12.93$\pm$0.60\\
2.328\_C & --224    & 13.24$\pm$0.17 &  14.40$\pm$0.11 & $<$13.2         & $<$13.2        & $<$12.0        & $<$12.3        & $<$12.8        \\
2.321    & --632    & 14.11$\pm$0.08 &  15.04$\pm$0.05 & $<$13.2         & $<$13.2        & $<$12.0        & $<$12.3        & $<$12.8        \\
2.298    & --2729   & 13.43$\pm$0.33 &  14.16$\pm$0.20 &14.28$\pm$0.20$^b$&$<$13.2        & $<$12.0        & 12.68$\pm$0.20 & 13.22$\pm$0.10\\
2.296    & --2913   & 13.79$\pm$0.16 &  14.72$\pm$0.25 & 13.43$\pm$0.40  & $^b$           & $<$12.0        & $<$12.3        & 12.68$\pm$0.21\\
\tableline
\end{tabular}
\normalsize

errors, upper and lower limits are 90\% confidence intervals;
$^{a}$km/s; $^b$ the z=2.296 \ion{C}{2}$^*$ line is completely blended with the z=2.298 \ion{C}{2} line;
$^c$ for system z=2.328\_A we measuered also a column density of \ion{Al}{3} of 13.66$\pm$0.30.

\end{table}

\begin{figure}[h]
\centering
\includegraphics[width=10cm]{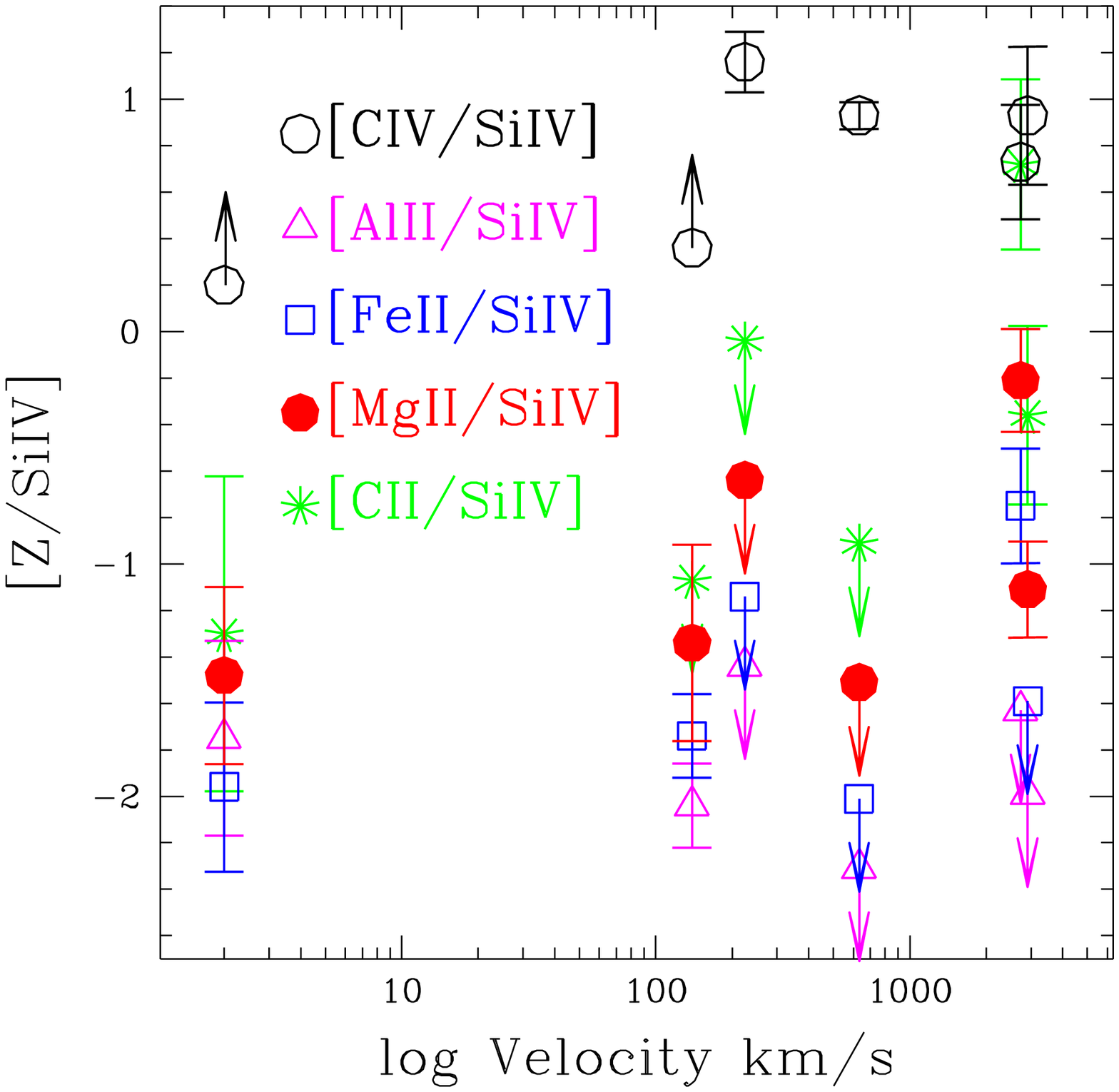}
\caption{The logarithmic ratio between the \ion{C}{4}, \ion{C}{2}, \ion{Fe}{2},
\ion{Al}{2} and \ion{Mg}{2} column densities to that of \ion{Si}{4}
for the six absorption systems as a function of the negative velocity
shift with respect to the redshift of the host galaxy.  }
\label{relden}
\end{figure}

The detection of both high and low ionization lines, feasible thanks
to the extremely wide spectral coverage achieved by UVES, allows us to
obtain constraints on the ionization status of the gas responsible for
the UV absorption, by comparing ion column density ratios with the
predictions of photoionization codes.  Unfortunately we were not able
to measure column densities of different ions of the same element,
except for \ion{C}{2}/\ion{C}{4} and \ion{Al}{2}/\ion{Al}{3} for
system z=2.328\_A, and \ion{C}{2}/\ion{C}{4} for system z=2.298.  In the
other cases we are forced to use ratios of column densities of
different ions of different elements in this analysis. These estimates
of the gas ionization parameter are therefore somewhat degenerate with
respect to relative element abundances.  We used the Grevesse \&
Anders (1989) meteoritic abundances with extensions by Grevesse et
al. (1993).

\begin{figure}
\vbox{
\hbox{
\psfig{figure=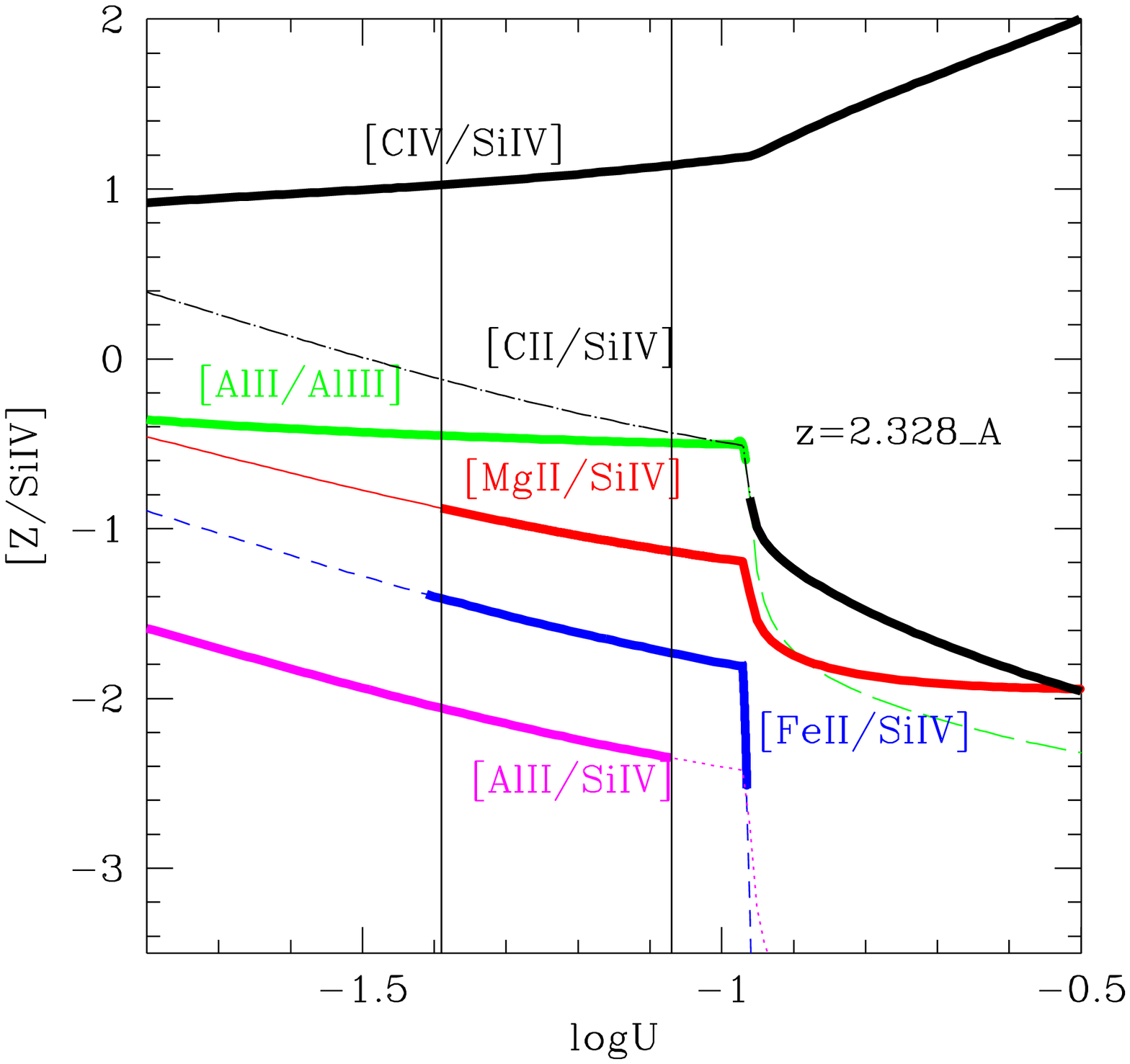,height=6.8cm,width=8cm}
\psfig{figure=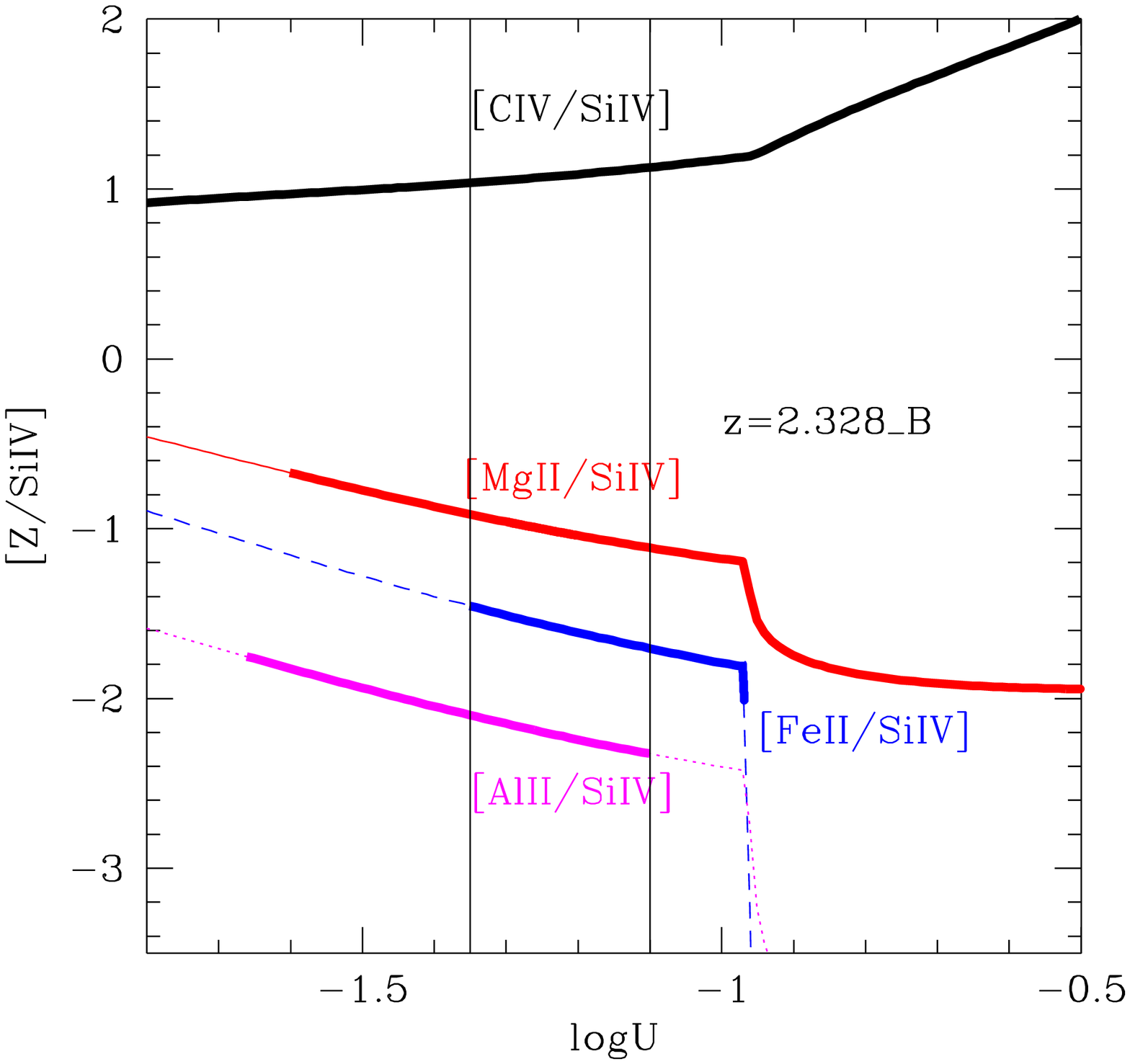,height=6.8cm,width=8cm}
}
\hbox{
\psfig{figure=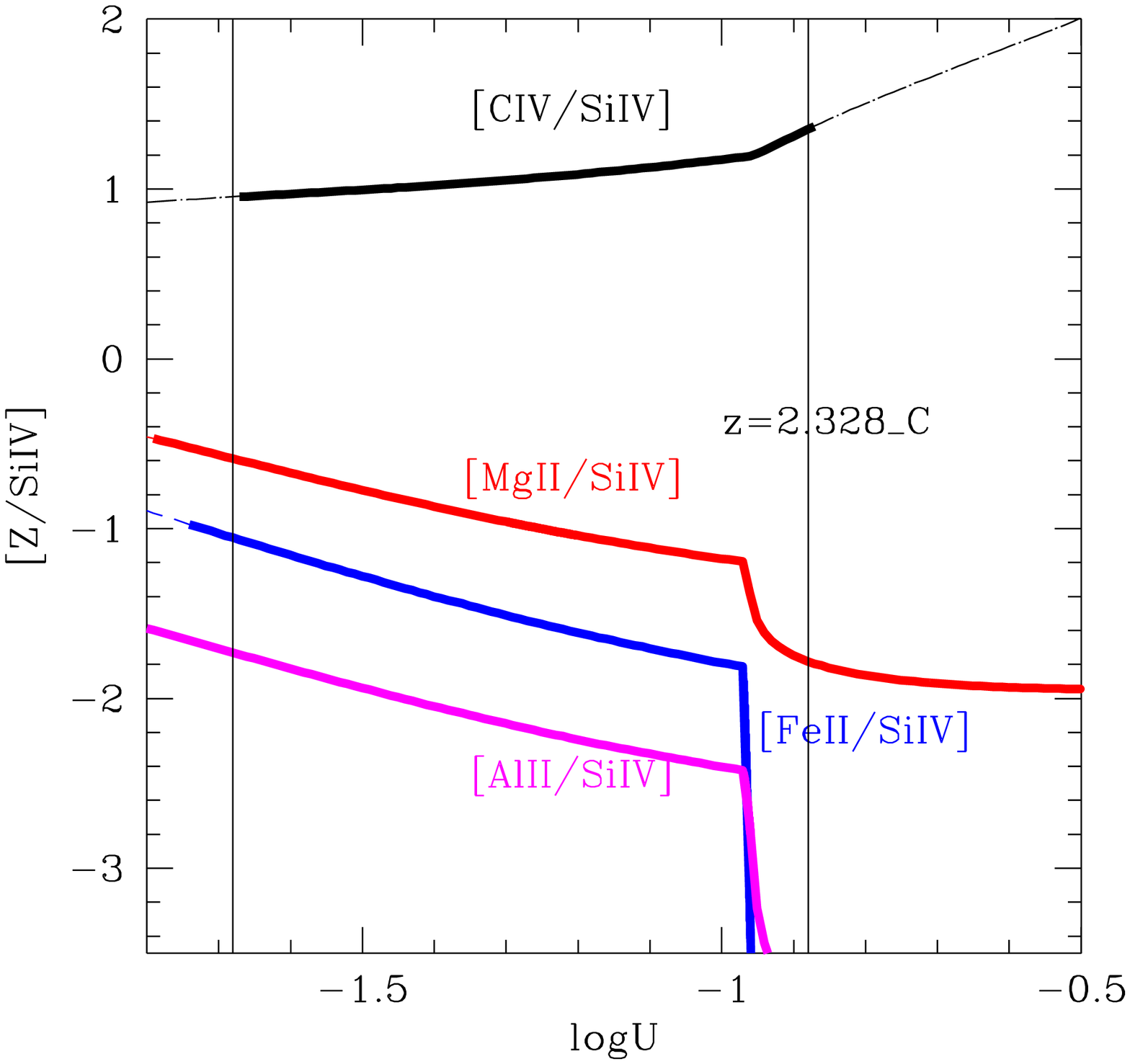,height=6.8cm,width=8cm}
\psfig{figure=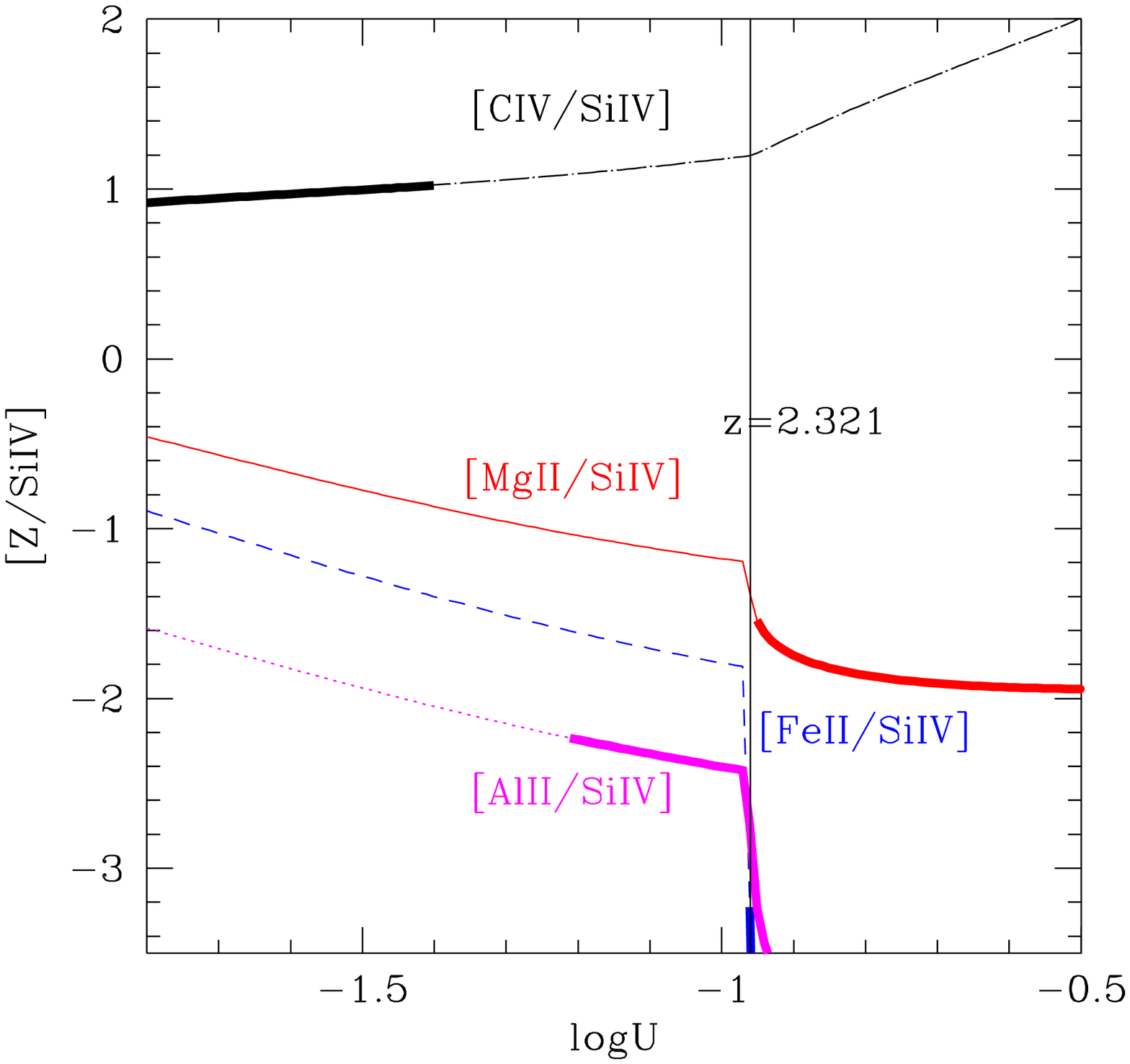,height=6.8cm,width=8cm}
}
\hbox{
\psfig{figure=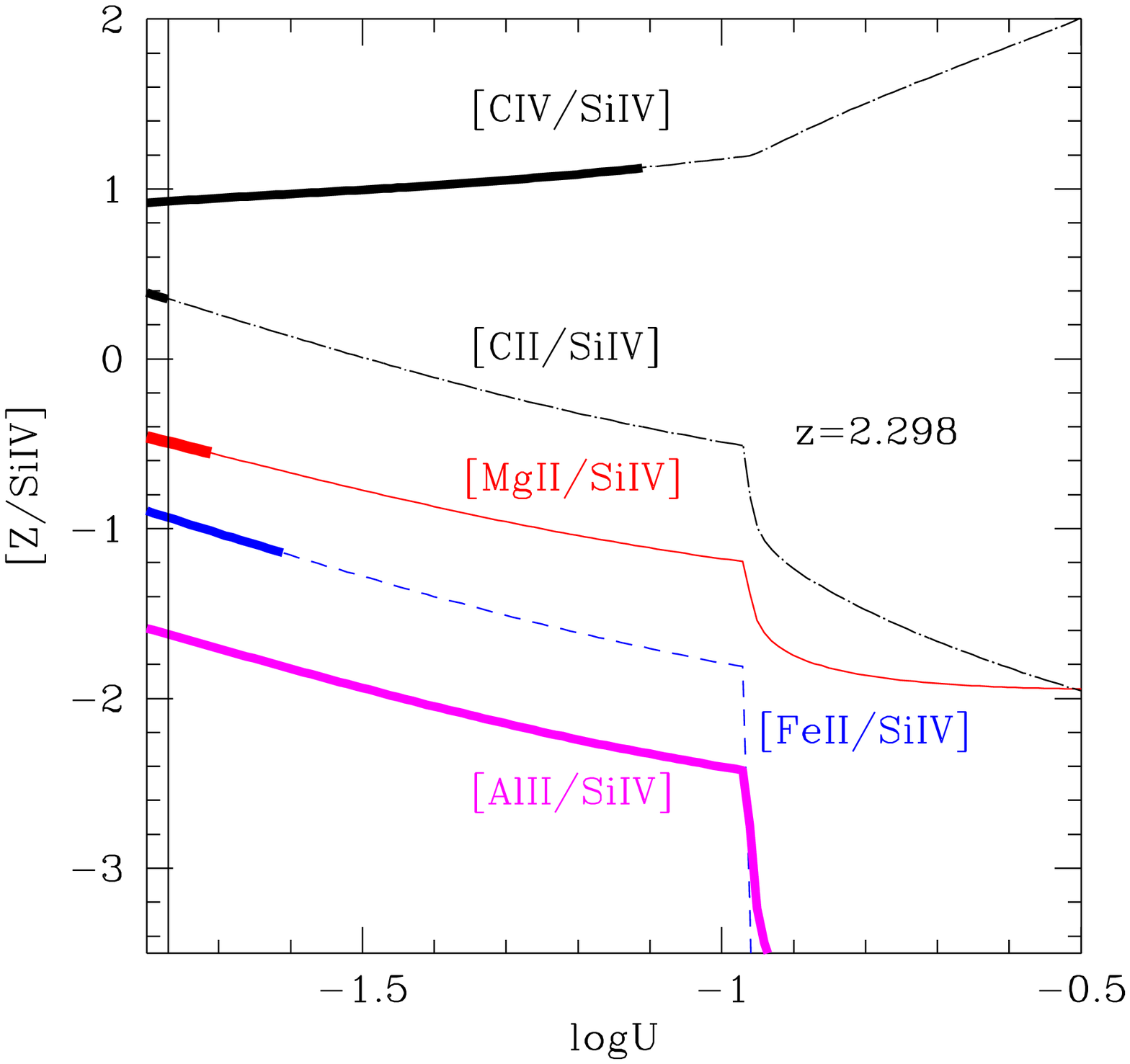,height=6.8cm,width=8cm}
\psfig{figure=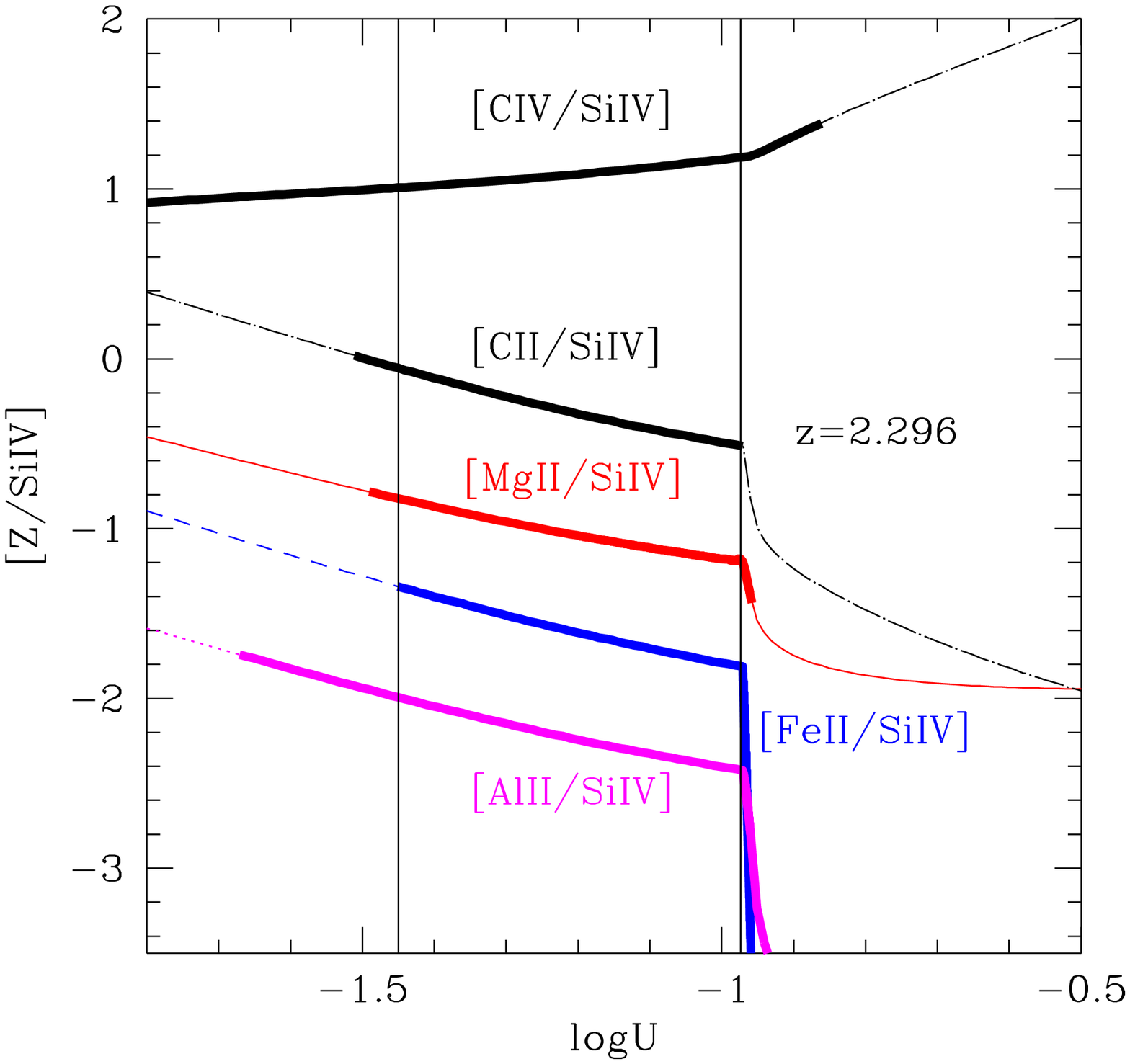,height=6.8cm,width=8cm}
}
}
\vspace{-0.3cm}
\caption{Ion ratios from CLOUDY models for the six absorption
systems, assuming a gas density of 1 cm$^{-3}$:
[\ion{Mg}{2}/\ion{Si}{4}] = solid red curves,
[\ion{Fe}{2}/\ion{Si}{4}]= dashed blue curves,
[\ion{Al}{2}/\ion{Si}{4}] = dotted magenta curves and
[\ion{C}{4}/\ion{Si}{4}] and [\ion{C}{2}/\ion{Si}{4}]= dot-dashed black
curves.  The line indicating the model values is thick when it is
consistent within 90\% with the measured ion ratios.  The vertical
solid lines mark the range of allowed U values in each panel.}

\label{ion2}
\end{figure}

\begin{figure}
\vbox{
\hbox{
\psfig{figure=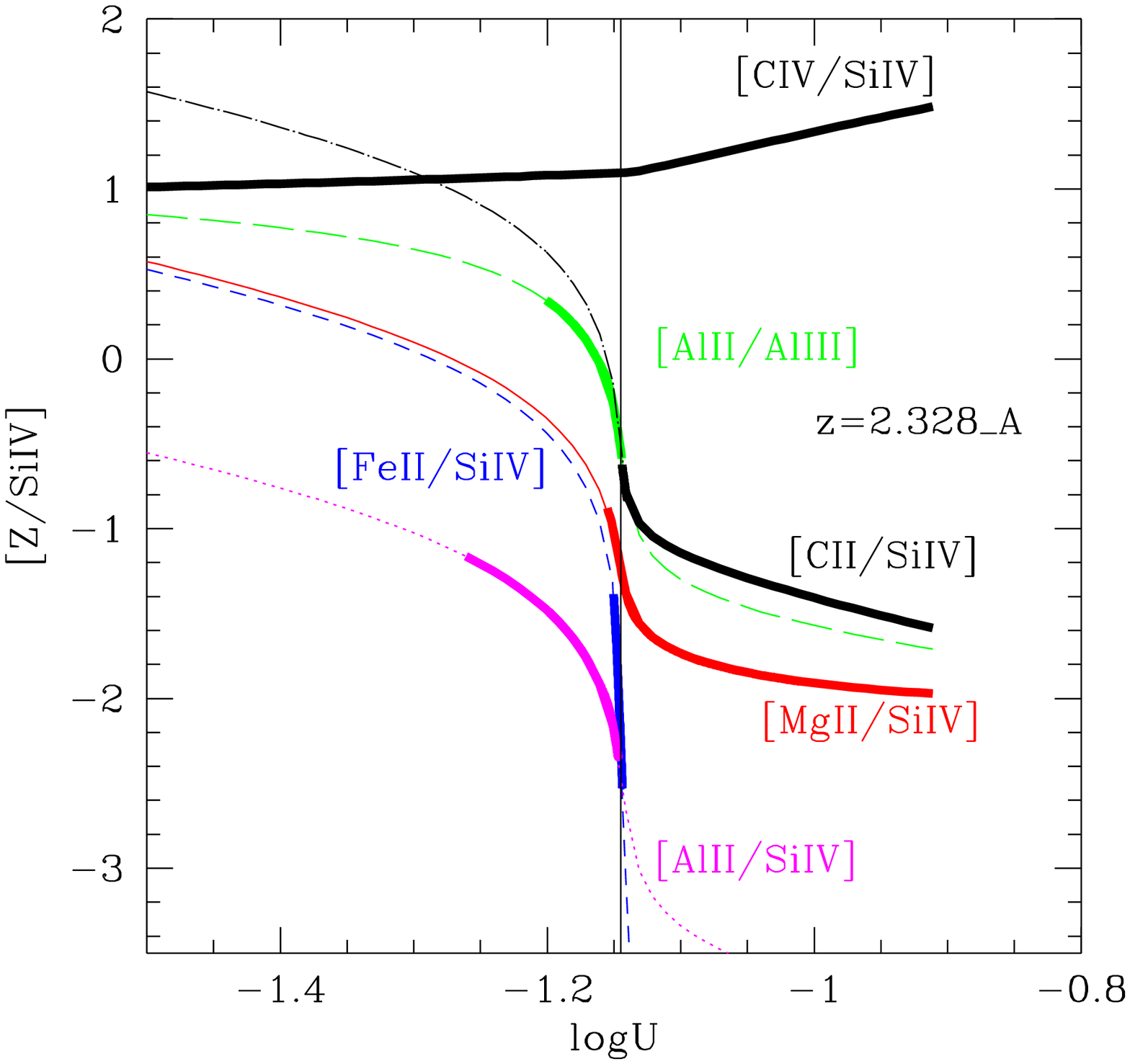,height=6.8cm,width=8cm}
\psfig{figure=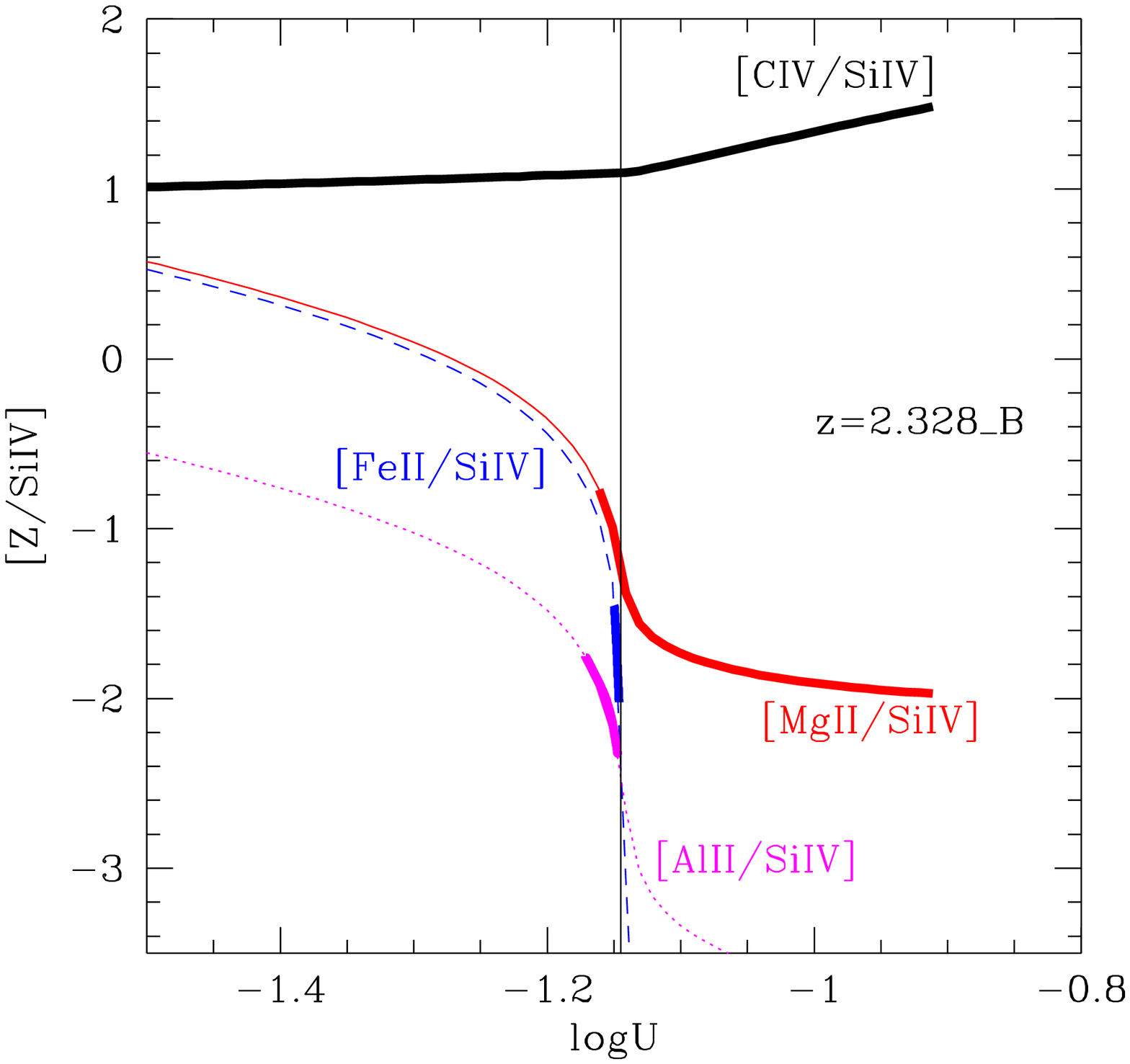,height=6.8cm,width=8cm}
}
\hbox{
\psfig{figure=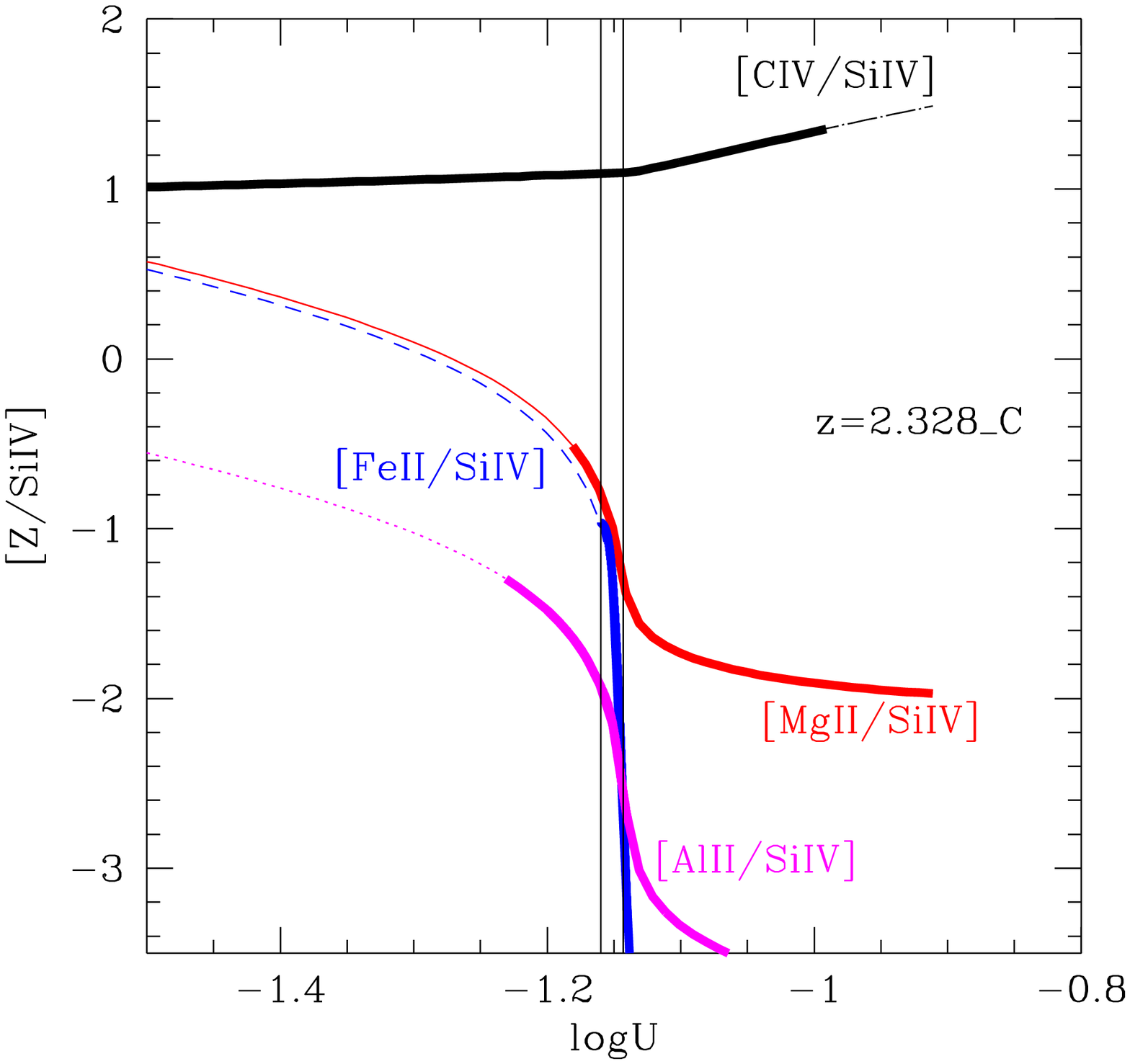,height=6.8cm,width=8cm}
\psfig{figure=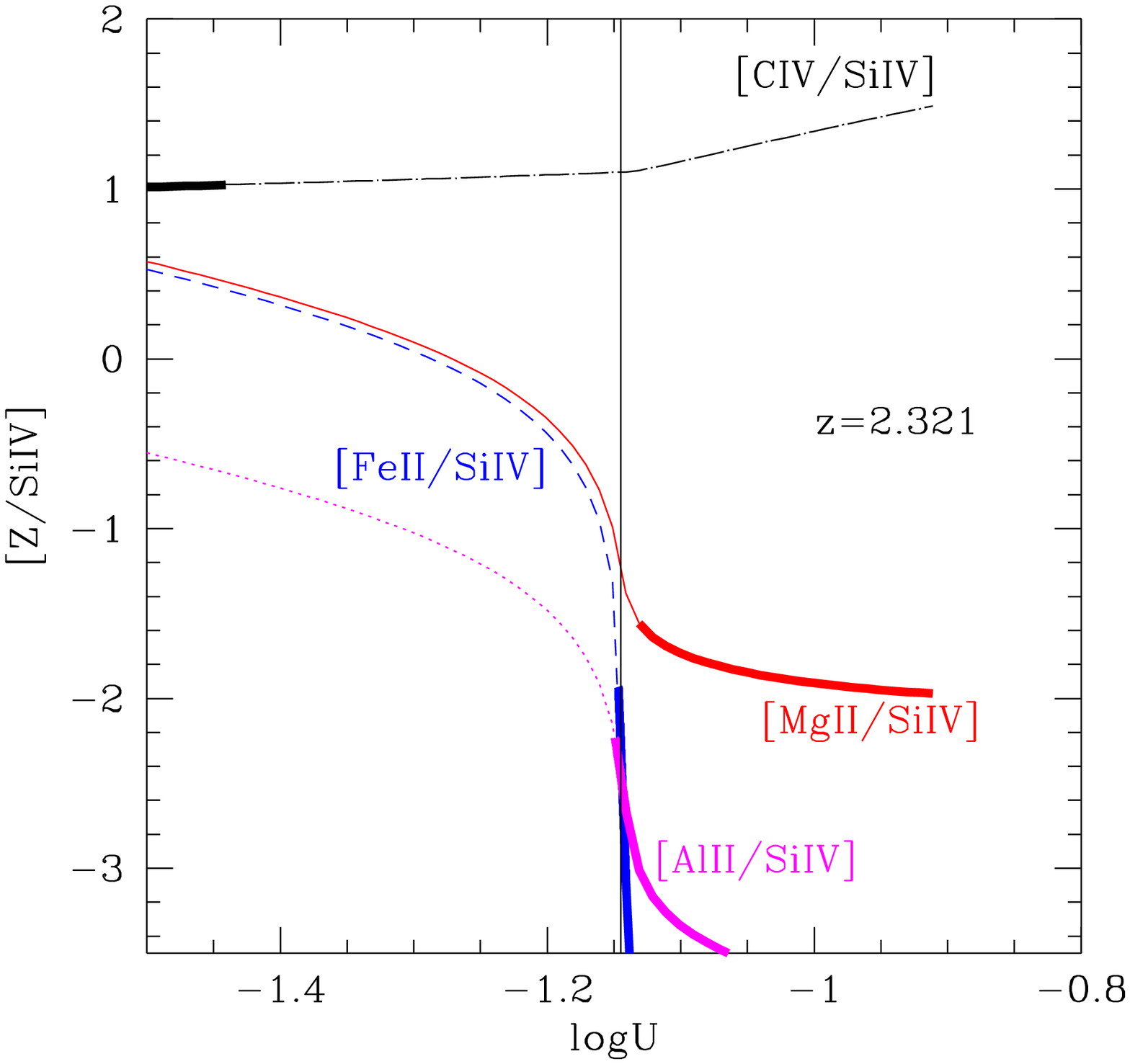,height=6.8cm,width=8cm}
}
\hbox{
\psfig{figure=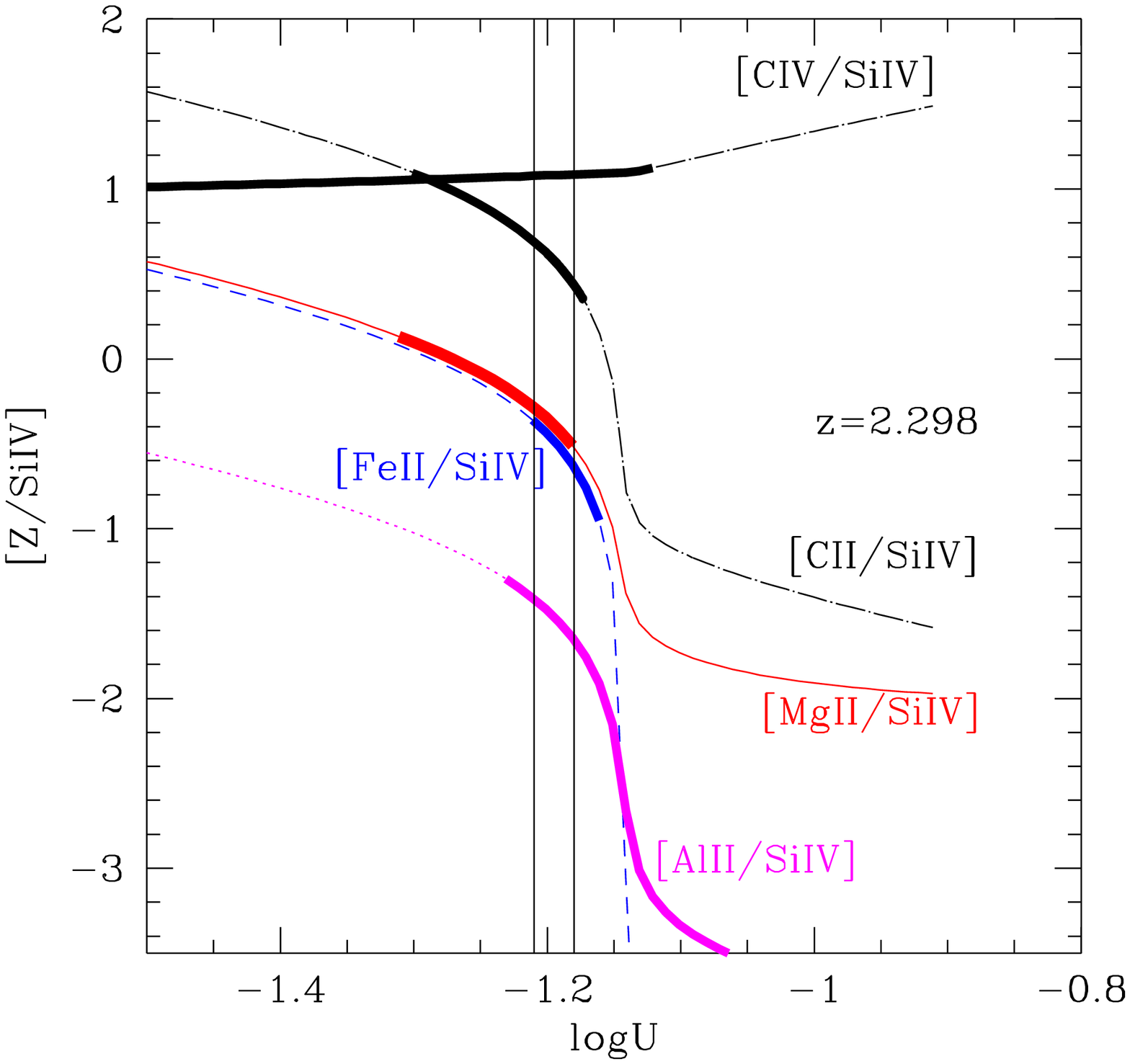,height=6.8cm,width=8cm}
\psfig{figure=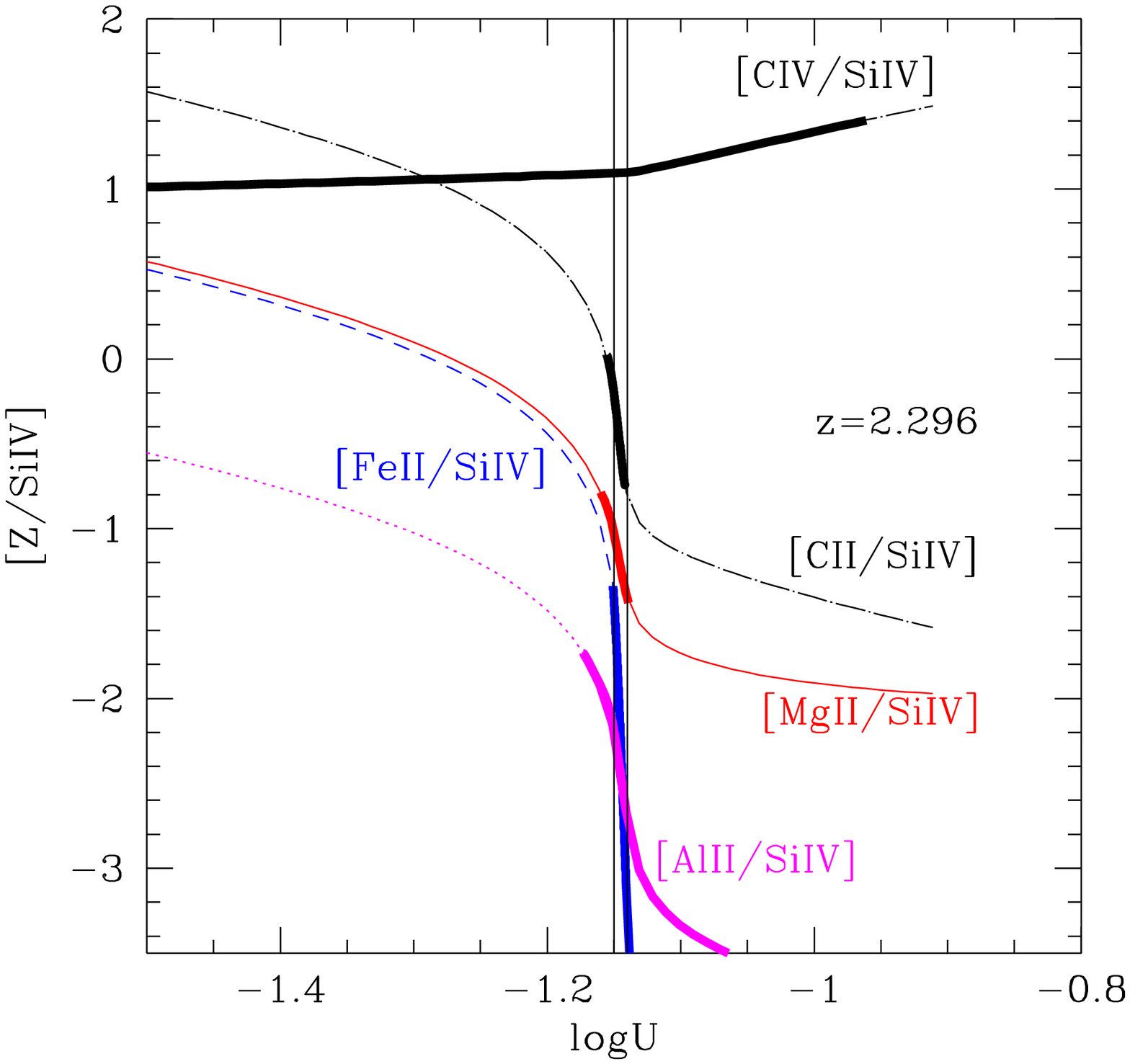,height=6.8cm,width=8cm}
}
}
\caption{ Ion ratios from CLOUDY photoionization models for the
six absorption systems assuming a gas density of 10$^8$ cm$^{-3}$.
Curves as in figure \ref{ion2}. The line indicating the model values
is thick when it is consistent within 90\% with the measured ion
ratios, as in figure \ref{ion2}.  The vertical solid lines mark the
allowed U values in each panel.}
\label{ion2_8}
\end{figure}

Figure \ref{relden} shows the logarithmic ratio between the
\ion{C}{4}, \ion{C}{2},\ion{Fe}{2}, \ion{Al}{2} and \ion{Mg}{2}
column densities and that of \ion{Si}{4} for the six absorption
systems as a function of the velocity shift with respect to the
redshift of the host galaxy. No large variation of the ion ratios are
seen for the six systems. We compared these line ratios to the
predictions obtained by simulating a gas cloud illuminated by an
ionizing continuum. We used Cloudy (vs 90.04, Ferland et al. 2002) to
build grids of photoionization models as a function of U, the
ionization parameter. U is defined as the ratio between the ionizing
photon density and the electron density of the gas. It is computed
assuming a constant density profile throughout the cloud, and a plane
parallel geometry.  We studied gas densities between 1 cm$^{-3}$ and
$10^{8}$ cm$^{-3}$.  The ionizing continuum was assumed to be a power
law, $F(E)=E^{-\Gamma}$ photons cm$^{-2}$ s$^{-1}$, with cutoffs at
low and high energies. The high energy cutoff is fixed at $10^{21}$Hz,
while we run the simulations for different low energy cutoffs, from
$10^{10}$Hz to $10^{14}$Hz.  We produced several sets of simulations
with $\Gamma$ in the range 1--2.  The ionizing continuum is constant
in time. GRBs are highly variable sources and the ionization structure
of the gas can be better studied using a time--dependent
photoionization code, such as those of Nicastro et al. (1999) and
Perna \& Lazzati (2002).  Nevertheless, our simpler approach is
instructive in identifying general trends.

\begin{figure}
\centering
\includegraphics[width=10cm]{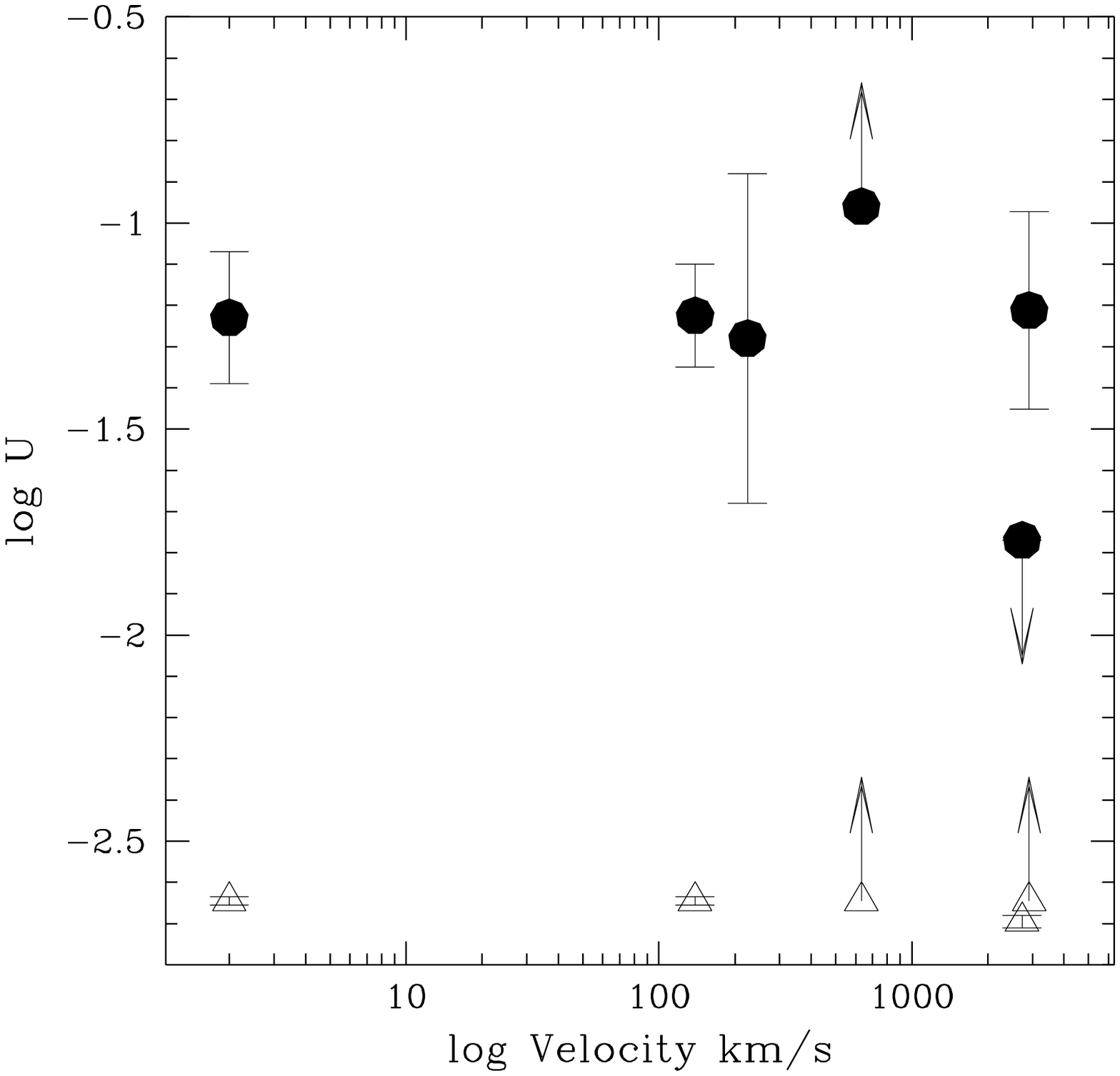}
\caption{The logarithm of the best fit ionization parameters from
figure \ref{ion2} (n=1 cm$^{-3}$ filled circles) and from figure 
\ref{ion2_8} (n=$10^8$ cm$^{-3}$, open
triangles, shifted by logU=-1.5), as a function of the velocity shift
with respect to the redshift of the host galaxy. Error-bars and upper
limits represent the purely statistical 90\% confidence intervals.}
\label{uv}
\end{figure}

The predicted photoionization equilibrium ion ratios of
[\ion{Mg}{2}/\ion{Si}{4}], [\ion{Fe}{2}/\ion{Si}{4}],
[\ion{Al}{2}/\ion{Si}{4}] and [\ion{C}{4}/\ion{Si}{4}] are shown in
figure \ref{ion2} for $\Gamma=2$ and n=1 cm$^{-3}$ and figure
\ref{ion2_8} for $\Gamma=2$ and n=$10^8$ cm$^{-3}$. For each of the
six redshift systems, thick segments superimposed on these curves
indicate our 90\% confidence determinations for these ratios. In
figure \ref{ion2} (for n=1 cm$^{-3}$) the [\ion{Mg}{2}/\ion{Si}{4}],
[\ion{Fe}{2}/\ion{Si}{4}] and [\ion{Al}{2}/\ion{Si}{4}] ion ratios for
the systems z=2.328\_A, z=2.328\_B, z=2.328\_C, and z=2.296 are all
consistent with the same ionization status.  For the system z=2.328\_A
we measured the column density of \ion{Al}{2} and \ion{Al}{3}. Also
their ratio is consistent with the ionization parameter inferred from
the above ion ratios, see figures \ref{ion2} and \ref{ion2_8}.  
For the same system we measured the column density of \ion{C}{2}. The
[\ion{C}{2}/\ion{Si}{4}] ion ratio indicates a ionization parameter
somewhat higher than that indicated by the other ion ratios for n=1
cm$^{-3}$ (see figure \ref{ion2}), while for n=$10^8$ cm$^{-3}$ the
ionization parameter is fully consistent with that indicated by the
other ion ratios (figure \ref{ion2_8}).  The system at z=2.298
shows a ionization parameter lower than that of system z=2.328\_A,
while the system at z=2.321 shows a higher ionization parameter.
However, all ionization parameters systems are consistent with a
remarkably narrow range: $10^{-1.7}<U<10^{-1}$, for n=1 cm$^{-3}$ and
$10^{-1.2}<U<10^{-1}$, for n=$10^8$ cm$^{-3}$ see figure \ref{uv},
which plots the ionization parameters as a function of the velocity
for the six systems. We note that in these ranges of U the
[\ion{Fe}{2}/\ion{Mg}{2}] ratio varies by less than 30\%, and the
value found for the three systems in which we detect \ion{Fe}{2} and
\ion{Mg}{2} lines ([\ion{Mg}{2}/\ion{Fe}{2}]$\approx0.4$) is
consistent with what is expected from meteoritic abundances (and as
was adopted for the CLOUDY calculations).  This value is also in the
range found for the Galactic ISM  0.1$<$[Mg/Fe]$<$0.8. It is
somewhat higher than that found by Savaglio et al. (2004) in a sample
of faint K--band selected galaxies at 1.4$<$z$<$2.0 
-0.84$<$[Mg/Fe]$<$0.13.  Assuming meteoritic abundances the hydrogen
column density of the densest system (z=2.328\_A) is
$\gs5\times10^{19}$ cm$^{-2}$, where the equality yields if all Si is
in the form of \ion{Si}{4}. This column is so low that would have
easily escaped detection in the Chandra X--ray spectrum (Butler et
al. 2003) and it is consistent with the upper limit given by
Moller et al. (2003).

Qualitatively similar results are drawn from the analysis of the
n=$10^8$ cm$^{-3}$ curves, although the range of acceptable ionization
parameters for the 6 systems is narrower, see figure \ref{ion2_8}.
Qualitatevely similar results were also obtained for the $\Gamma=1$
models.

\subsubsection{The z=2.328\_A system}

For this system we detect the \ion{C}{2}$^*\lambda$1335 fine structure
line. Unfortunately, the \ion{C}{2}$\lambda$1334 ground state line is
barely visible (with a signal to noise of just 2).  The presence
of a strong \ion{C}{2}$^*$ line suggests either a high gas density or a
strong radiation field (Srianand \& Petitjean 2000, 2001, Silva \&
Viegas 2002). \ion{C}{2}$^*$ and \ion{Si}{2}$^*$ fine structure lines
have been detected in two other GRBs: GRB 020813 (Savaglio \& Fall 2004)
and GRB 030323 (Vreeswijk et al. 2004). Their ratio can be used to
constrain the gas density or the radiation field (if both 
lines are detected one may solve for both variables, as shown by
Srianand \& Petitjean 2000), so their detection in GRB spectra looks
very promising to constrain the physical properties of the
absorbing gas.

For this system the \ion{C}{4} line is strongly saturated and so it
cannot be used together with the \ion{C}{2} to constrain the gas
ionization status.  As discussed in the previous section, within
rather large uncertainties, the [\ion{C}{2}/\ion{Si}{2}] would suggest
for n=1 cm$^{-3}$ a ionization parameter somewhat higher than that
indicated by all other line ratios, assuming meteoritic
abundances. This may be explained by a slight under-abundance of C
with respect to Si in this system.  A similar conclusion applies to
system z=2.321, see below.

The \ion{C}{4}$\lambda1548$ lines of the z=2.328 A,B and C systems
are blended with the  \ion{C}{4}$\lambda1550$ of the 
z=2.321 system, suggesting the presence of line locking, 
as first reported by Savaglio et al. (2002). Line locking
is usually interpreted as the signature of line radiation
pressure acceleration (see e.g. Foltz et al. 1987, 
Srianand \& Petitjean 2000).

\subsubsection{The z=2.321 system}

For this system (as for z=2.328\_C) we were able to detect \ion{C}{4}
and \ion{Si}{4} absorption lines only. The upper limits on the
\ion{Mg}{2} and \ion{Fe}{2} transition allow us to put a significant
lower limit on the ionization parameter of U$\gs10^{-1}$ for this
system for a gas density of 1 cm$^{-3}$ and U$\gs10^{-1.2}$ for a gas
density of 10$^8$ cm$^{-3}$. For n=1, this is higher than the values
for the other systems at a confidence level better than 90\%. 
The [\ion{C}{4}/\ion{Si}{4}] ratio is formally inconsistent with this
higher ionization parameter, but the disagreement is marginal, taking
into account that the \ion{C}{4}/\ion{Si}{4} curves in figures
\ref{ion2} and \ref{ion2_8} are very flat, and therefore small
differences in the obsterved \ion{C}{4}/\ion{Si}{4} ratio would
translate in large differences in the ionization parameter. Also in
this case the disagreement may be explained by a slight under-abundance
of C with respect to Si.

\subsubsection{The z=2.296 and z=2.298 systems}

Given the relatively high velocity shift of about 3000 km/s between
the z=2.296 and z=2.298 systems and the host galaxy redshift, there is
the possibility, at least in principle, that these systems are not
associated with the GRB host galaxy but rather are intervening
systems.  We evaluated the probability of finding intervening \ion{C}{4}
systems within 3000 km/s from the GRB host galaxy, from the density
distribution of \ion{C}{4} systems given in D'Odorico et al. (1998) and
Petitjean \& Bergeron (1994).  The probability of finding by chance
two systems with column densities equal or larger than those given in
Table 5 for these two lowest redshift systems is $1.2\times10^{-5}$,
while the probability for only one of these systems is 0.1\% for the
z=2.296 system and 0.5\% for the z=2.298 system. 

The z=2.298 system has a ionization parameter $<10^{-1.7}$ at the
90\% confidence level for a gas density of 1 cm$^{-3}$ (around
$10^{-1.2}$ for a gas density of 10$^8$ cm$^{-3}$).  This is the
lowest of the ionization parameters found in the 6 systems.
Furthermore, for this system we detect a rather strong
\ion{C}{2}$\lambda$1334 ground state line but not the
\ion{C}{2}$^*\lambda$1335 high excitation line, see figure \ref{z2},
again indicating a lower radiation field (or a lower gas density).

The observed wavelength of the  z=2.298 system \ion{C}{2}$\lambda$1334 
line coincides with the wavelength of the z=2.296 \ion{C}{2}$^*\lambda$1335
line (see Table 3). This may be a further indication of line locking.



\subsubsection{Intervening systems}

Two main intervening systems are present along the line
of sight to GRB 021004, one around z=1.60 and the other around z=1.38
(see eg. Mirabal et al. 2003, Moller et al 2003). The UVES high
resolution spectrum of GRB 021004 allows an accurate study of these systems,
which are split in several components (see Table 3).

The system at z=1.38 consists at least 3 components at z=1.3807
(z=1.38\_A), z=1.3802 (z=1.38\_B), z=1.3795 (z=1.38\_C). The latter
system is also split in several parts in the range 1.3793-1.3798. The
total velocity range spanned by the system is $\sim180$ km/s. We
detected the following lines belonging to this system:
\ion{Si}{2}$\lambda$1808, \ion{Al}{3}$\lambda$1854,
\ion{Fe}{2}$\lambda$2344,\ion{Fe}{2}$\lambda$2374,
\ion{Fe}{2}$\lambda$2586,\ion{Mg}{2}$\lambda \lambda$2796,2803, and
\ion{Mg}{1}$\lambda$2852.

The system at z=1.60 is made by at least 4 components at z=1.6028
(z=1.60\_A), z=1.6024 (z=1.60\_B), z=1.6019 (z=1.60\_C) and z=1.6014
(z=1.60\_D).  The total velocity range spanned by the system is of
$\sim160$ km/s. For this system we detected the following lines:
\ion{Al}{2}$\lambda$1670,
\ion{Fe}{2}$\lambda$2344,\ion{Fe}{2}$\lambda$2374,
\ion{Fe}{2}$\lambda$2382,
\ion{Fe}{2}$\lambda$2586,\ion{Mn}{2}$\lambda$2594,
\ion{Fe}{2}$\lambda$2600,\ion{Mn}{2}$\lambda$2606, \ion{Mg}{2}$\lambda
\lambda$2796,2803, and \ion{Mg}{1}$\lambda$2852.

We estimated the column densities of the ions of the
intervening systems using the same fitting procedure used
for the six systems associated with the GRB host galaxy.
They are given in Table 6

\begin{table}
\caption{\bf Log. ion column densities in cm$^{-2}$ of the GRB 021004 
intervening systems}
\footnotesize
\begin{tabular}{lccccc}
\tableline\tableline
System &  v. shift$^a$ &\ion{Fe}{2}&  \ion{Mg}{2}    & \ion{Si}{2} & \ion{Al}{3}\\
\hline
z=1.38\_A & 0     & 15.68$\pm$0.16 & 13.23$\pm$0.07  & 15.12$\pm$0.20 & 13.02$\pm$0.20 \\       
z=1.38\_B & --63  &      --        & 12.92$\pm$0.32  &   --           &  --            \\       
z=1.38\_C & --150 & 14.75$\pm$0.40 & 15.20$\pm$0.70  &   --           &  --            \\       
\hline
System &  v. shift$^a$ &\ion{Fe}{2}&  \ion{Mg}{1}    & \ion{Mg}{2} & \ion{Al}{2}\\
\hline
z=1.60\_A &  0    & 13.69$\pm$0.16 & 11.33$\pm$0.26  & 13.54$\pm$0.21 & 12.66$\pm$0.27 \\       
z=1.60\_B & --46  & 12.63$\pm$0.16 & 11.64$\pm$0.18  & 12.95$\pm$0.10 & --             \\       
z=1.60\_C & --104 & 14.35$\pm$0.18 & 12.80$\pm$0.27  & 15.45$\pm$0.80 & 13.24$\pm$0.50 \\       
z=1.60\_D & --160 & 13.57$\pm$0.10 & 11.82$\pm$0.14  & 13.47$\pm$0.06 & 12.58$\pm$0.25 \\       
\tableline
\end{tabular}
\normalsize

errors, upper and lower limits are 90\% confidence intervals;
$^{a}$km/s

\end{table}

\begin{table}
\caption{\bf GRB 020813, Logarithmic ion column densities in cm$^{-2}$}
\footnotesize
\begin{tabular}{lccc}
\tableline\tableline
System & velocity shift$^{a}$ & \ion{Fe}{2} & \ion{Mg}{2}  \\
\hline
z=1.255\_A & 0.       & 15.20$\pm$0.29 &  $>16.0$   \\
z=1.255\_B & --106    & 15.47$\pm$0.27 & $>15.5$  \\
z=1.255\_C & --306    & 13.58$\pm$0.10 &  13.79$\pm$0.21 \\
z=1.2234   & --4204   & 13.96$\pm$0.18 & $>16.0$  \\              
\tableline
\end{tabular}
\normalsize

$^{a}$km/s

\end{table}

\subsection{GRB 020813}

\begin{figure}
\centerline{ 
\psfig{figure=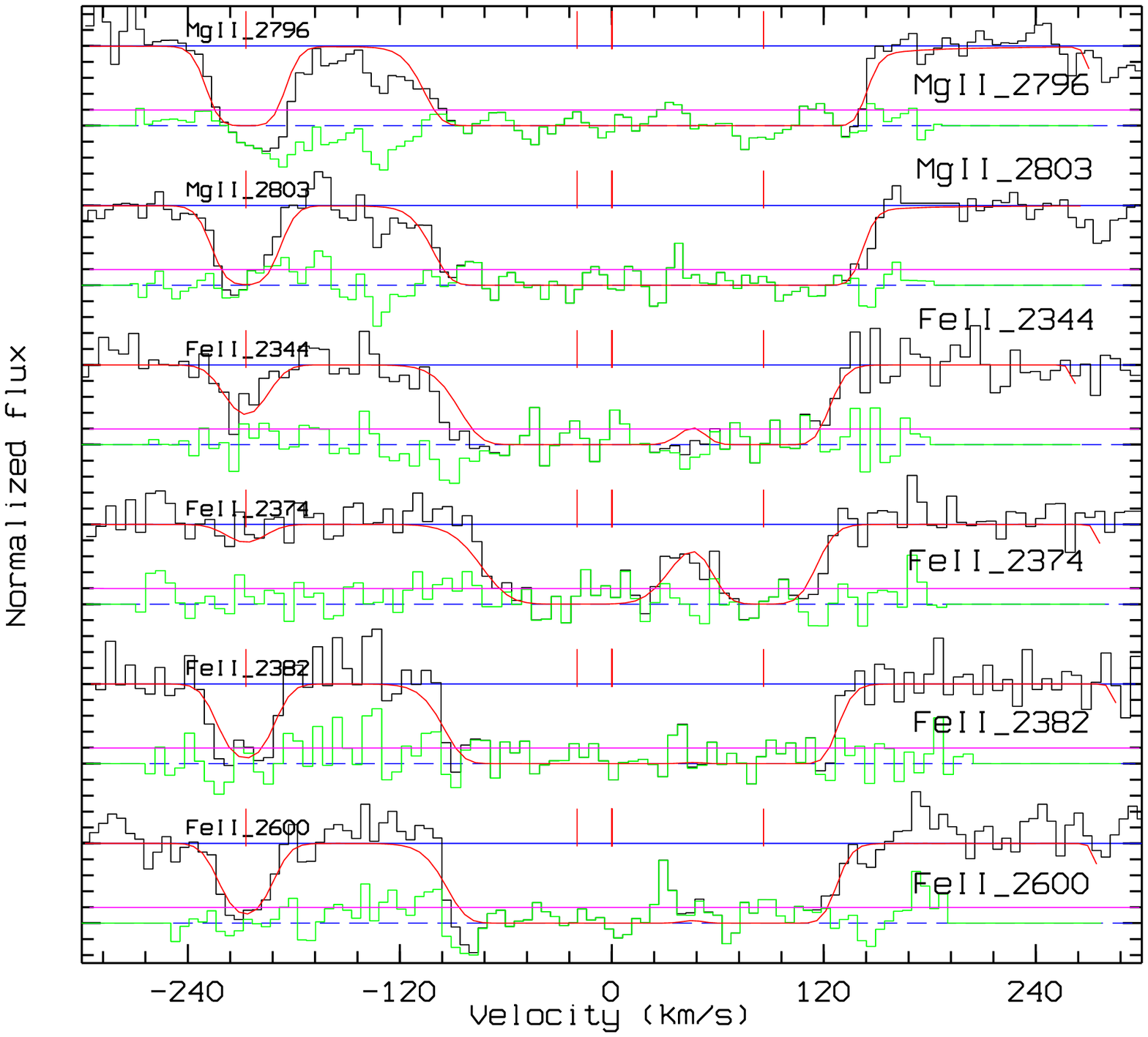,width=10cm,angle=-90}
}
\caption{UVES spectrum of GRB 020813 near the \ion{Fe}{2}$\lambda$2600,
\ion{Fe}{2}$\lambda$2382, \ion{Fe}{2}$\lambda$2374,
\ion{Fe}{2}$\lambda$2344, \ion{Mg}{2}$\lambda$2803 and
\ion{Mg}{2}$\lambda$2796 lines for the three absorption systems, in
velocity space, along with the best fit model, solid line and
residuals. The zero of the velocity scale refers to z=1.2545}
\label{grb020813}
\end{figure}

For this GRB we considered the absorption systems at the following
velocities with respect to the redshift of the host galaxy, which was
assumed to be 1.255 (Barth et al. 2003): v=0 km/s (system A in figure
\ref{grb020813}), v=--106 km/s (system B) and v=--306 km/s (system C).
Note that the system A and system B lines are strongly blended with
the exception of the \ion{Fe}{2}2374 line. We also consider the
system at z=1.2234.

We fitted simultaneously the 6 lines of the 3 absorption systems in
figure \ref{grb020813}. Table 7 gives for the 3 systems the best fit
\ion{Fe}{2} and \ion{Mg}{2} abundances along with the velocity shift
of each system with respect to the redshift of the host galaxy.
\ion{Mg}{2} lines of systems A and B are strongly saturated and
therefore their column density estimates are more uncertain.
Unfortunately the UVES spectrum covers a wavelength range much smaller
than the Keck LRIS spectrum and several of the lines studied by Barth
et al. (2003) and Savaglio \& Fall (2004), in particular \ion{Zn}{2},
\ion{Cr}{2}, \ion{Si}{2}, are not accessible.

\begin{figure}
\centerline{ 
\psfig{figure=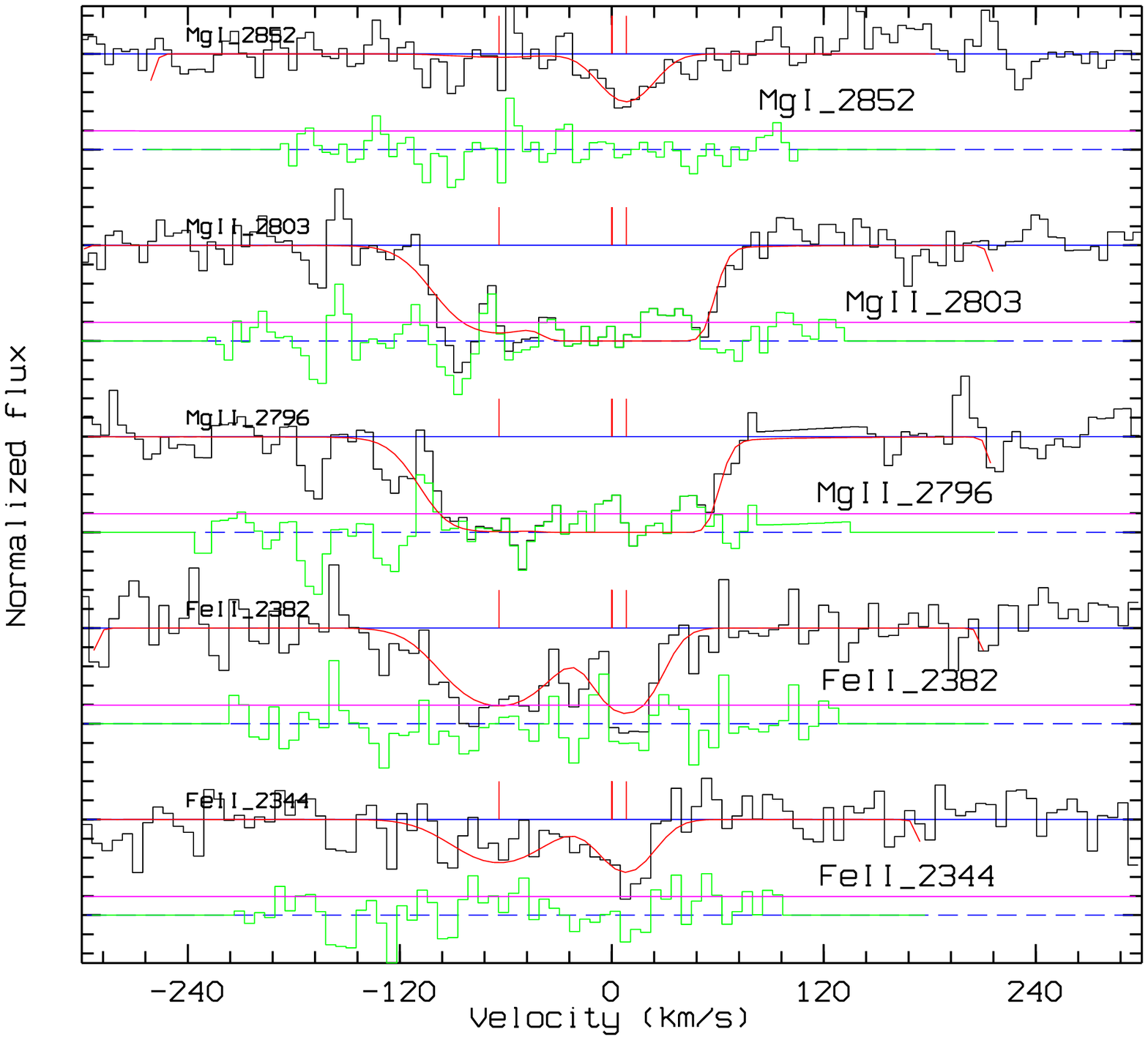,width=10cm,angle=-90}
}
\caption{UVES spectrum of GRB 020813 near the 
\ion{Fe}{2}$\lambda$2382, 
\ion{Fe}{2}$\lambda$2344, \ion{Mg}{2}$\lambda$2803,
\ion{Mg}{2}$\lambda$2796 and \ion{Mg}{1}$\lambda$2852 lines 
for the z=1.2234 absorption systems, in
velocity space, along with the best fit model, solid line and
residuals. The zero of the velocity scale refers to z=1.2234}
\label{grb020813_12234}
\end{figure}

We performed similar fits to the 5 lines associated with the 
z=1.2234 system, split in two components. 
The results are in figure \ref{grb020813_12234} and in Table 7.

Unfortunately in this case the redshift of the GRB is not high enough
to have strong high ionization lines in the spectrum, and therefore we
do not have a direct way to constrain the ionization status of the gas
responsible for the UV absorption.

The [\ion{Fe}{2}/\ion{Mg}{2}] ratio is consistent with a constant
value in the four systems.  This ratio is also consistent with
meteoritic abundances, assuming that the ionization parameter is in
the range $10^{-2.5} - 10^{-1}$.

\subsection{The z=1.2234 system}

The system at z=1.2234 is shifted by about 4200 km/s from the redshift
of the host galaxy. The probability to find an \ion{Mg}{2} intervening
system with $W_{\lambda 2796.35}=1.3\pm0.03\AA\ $ (see Table 7) within
this velocity range is $\ls1\%$, using the distribution of \ion{Mg}{2}
systems given by Steidel \& Sargent (1992).  Although not as
conclusive as in the case of the z=2.296 and z=2.298 systems in the
spectrum of GRB 021004, we remark that this probability is rather
small, and that the velocity shift with respect to the redshift of the
host galaxy is intriguingly similar to that of the GRB 021004 systems.

\section{Discussion}

GRB 021004 shows absorption systems which span $\sim3000$~km/s in
velocity towards the observer (no receding system is detected). The
systems are most likely local to the GRB since the probability of
finding two or more different absorption systems along the line of
sight by random fluctuations is negligible. 

It has been suggested that radiative acceleration by the prompt GRB
emission may be responsible for the high detected speeds (Schaefer et
al. 2003). However, this seems unlikely since low ionization ions such
\ion{Fe}{2} and \ion{Mg}{2} are present in the high velocity absorber.
The radiative acceleration is dominated by the GRB and early afterglow
radiation. It is therefore unlikely that recombination can be so rapid
to have any influence on the acceleration of the absorbing
material. In the photoionization phase each nucleon receives an amount
of energy comparable to the ionization potential of the $K$--shell
electron. For a solar metallicity plasma the acceleration is regulated
by H atoms and, at the end of the photoionization, the absorbing
material has an outward velocity of:
\begin{equation}
v_{\rm{ion}}\sim\sqrt{\frac{h\nu_{\rm{ion}}}{m_p}}\sim50\quad{\rm km/s}
\end{equation}
Once the gas is fully ionized, acceleration proceeds due to radiation
pressure on free electrons through Inverse Compton (IC)
interactions. For a burst of isotropic equivalent energy
$E_{\rm{iso}}$ the IC acceleration is obtained by momentum
conservation:
\begin{equation}
v_{\rm{IC}}=\frac{E_{\rm{iso}}\,\sigma_T}{
4\pi\,R^2\,m_p\,c} \sim 0.6\,R_{18}^{-2}\quad{\rm km/s}
\end{equation}
we conclude that at the minimum distance of the absorbing medium
$R\sim10^{18}$~cm (Lazzati et al. 2002; Heyl \& Perna 2003; in order
to be ahead of the fireball at the time at which absorption is
detected) radiative acceleration is unable to propel the absorber to
an outflowing speed comparable to the value(s) observed.

Possible explanations for the large outflowing speed are either a
supernova exploded several years prior to the GRB (Vietri \& Stella
1998) or a high velocity wind from the progenitor Wolf--Rayet (WR)
star (MacFadyen \& Woosley 1999; Schaefer et al. 2003; Mirabal et
al. 2003). The large velocity spread with similar ionization parameter
is however difficult to account for in a SNR scenario, where values
around a typical expansion velocity would be expected. On the other
hand, WR winds are known to be clumpy and velocities up to
$\sim4000$~km/s were detected from P--Cygni profiles (Niedzlieski \&
Sk\'orzy\'nski 2002).

In this work we derived physical parameters of the absorbers under the
assumption of equilibrium conditions which, $\sim0.5$~days after the
GRB explosion, are attained only if the electron density in the
absorber is $n\sim 10^{7-8}$~cm$^{-3}$. Assuming equilibrium
conditions, the total column density of the absorber can be computed
from the ion column density corrected for the ionization
fraction. Consider the \ion{Si}{4} line in the A system. The density
is given by:
\begin{equation}
n=\frac{N_{\rm{Si}}}{A_{\rm{Si}}\,R\,\eta}
\sim\frac{10^{15.3}}{4\times10^{-5}\,10^{18}\,R_{18}\,10^{-6}\eta_{-6}}
\approx5\times10^7 \, R_{18}^{-1}\, \eta_{-6}^{-1} \qquad {\rm cm}^{-3}
\end{equation}
where $R$ is the fireball radius at the time the lines were observed,
$A_{\rm{Si}}$ is the Si abundance
and $\eta$ is the ratio of the width of the absorbing shell over its
radius. The ionization parameters derived above and implications
discussed here are therefore relevant only in the case of an extremely
clumpy wind or SNR.

The photoionization results of CLOUDY yield an ionization parameter
constrained in a relatively small range $10^{-1.7}<U<10^{-1}$. In a
single explosion GRB model, the ionization parameter scales with the
square of the outflow velocity, since the absorber's distance scales
with velocity as well. Even though the ionization parameter depends on
the electron density, it seems unlikely that density variations are
such as to compensate for the large velocity difference. In a wind
environment, on the other hand, the ionization parameter is constant
since the photon density and particle density scale with the same
power of the distance. The relatively small variations in the inferred
$U$, which do not show any clear trend with velocity, can therefore be
interpreted as density fluctuations on top of a regular $R^{-2}$ wind
density profile (as already discussed by e.g.  Schaefer et al. 2003
and Mirabal et al. 2003).

Finally, we note that the \ion{Fe}{2} and \ion{Mg}{2} column densities
found in GRB 020813 are 10--100 times higher than those in GRB 021004.
This is likely due to a much higher ionization of the gas in the
latter case, rather than a large difference in the total absorbing
column of gas or to highly non solar metal abundances (note that in
both cases the [\ion{Fe}{2}/\ion{Mg}{2}] ratio is consistent with
meteoritic abundances).  We also note the similarity between the
velocity shift between the z=1.2234 system and the GRB 020813 host
galaxy and the shift of the z=2.296 and z=2.298 systems with respect
to the GRB 021004 host galaxy. Although the probability for a chance
occurrence of the z=1.2234 system is not as low as in the cases of the
two shifted systems in GRB 021004, and that the z=1.2234 ionization
status is probably much lower than that of the z=2.296 and z=2.298
systems, this similarity might suggest a common scenario for the two
GRBs.

\section{Conclusions}

One of the straightforward results of our UVES high resolution
observations of two GRB afterglows is that the ISM of the host
galaxies is complex, and many components contribute to each main
absorption systems. These components span a total velocity range of up
to about 3000 km/s. Several narrow components are resolved down to a
width of a few tens of km/s.  The UVES wide band coverage allowed us
to investigate simultaneously both high ionization lines such as
\ion{C}{4} and \ion{Si}{4}, and low ionization lines such as
\ion{Mg}{2} and \ion{Fe}{2} in GRB 021004. This allowed us to constrain
the ionization parameter of the gas of the differerent absorption
systems. Combined with photoionization results obtained with CLOUDY
the ionization parameters appear to lie in a relatively small range,
with no clear trend with the system velocity. This can be interpreted
as density fluctuations on top of a regular $R^{-2}$ wind density
profile. The [\ion{Mg}{2}/\ion{Fe}{2}] ratio of $\approx0.4$ found for
systems z=2.328\_A, z=2.328\_B, and z=2.298 is consistent with what is
expected from meteoritic abundances. Assuming these abundances, the
measured \ion{Si}{4} column density for system z=2.328\_A implies a
lower limit of $\gs5\times10^{19}$ cm$^{-2}$ to the system total
hydrogen column density.

Indeed, our study shows that rapid reaction to the GRB triggers, high
resolution and wide spectral coverage are the key ingredients to study
GRB host galaxies. Today observations of this kind are still difficult
and rather episodic. However, they should become routine after the
launch of the {\it Swift} satellite, also thanks to the development of
a dedicated Rapid Response observing Mode for the VLT telescopes.
This mode will make possible automatic follow--up of GRBs (or other
transient events) with response times in the 10m--1hr range, therefore
helping in gathering spectra of unprecedented quality of medium to
high redshift GRB host galaxies.

\acknowledgments 
We acknowledge support from contract ASI/I/R/390/02
and MIUR grant Cofin--2003--41.  We thank Fabrizio Nicastro and
Emanuele Giallongo for usefull discussions, Sandra Savaglio for her
early work on this program, and an anonymous referee for comments
that helped to improve the presentation.

\end{document}